\documentclass{aa}

\usepackage{graphicx}
\usepackage{txfonts}
\begin{document} 

\title{Peering through the holes: the far-UV color of star-forming galaxies at $z\sim 3-4$ 
and the escaping fraction of ionizing radiation}

   \author{E. Vanzella \inst{\ref{Bol}},
           S. de Barros \inst{\ref{Bol}},
           M. Castellano \inst{\ref{Rom}},
           A. Grazian \inst{\ref{Rom}},
           A. K. Inoue \inst{\ref{osaka}}\thanks{Visiting Scholar at Department of Astronomy and Astrophysics,
University of California Santa Cruz, 1156 High Street, Santa Cruz, CA 95064, USA},
           D. Schaerer \inst{\ref{geneva},\ref{tolo}},
           L. Guaita, \inst{\ref{Rom}},
           G. Zamorani \inst{\ref{Bol}},\\
           M. Giavalisco \inst{\ref{UMASS}},
           B. Siana \inst{\ref{river}},
           L. Pentericci \inst{\ref{Rom}},
           E. Giallongo \inst{\ref{Rom}},
           A. Fontana \inst{\ref{Rom}},
           and C. Vignali \inst{\ref{unibo}}
          }
\institute{
INAF - Osservatorio Astronomico di Bologna, via Ranzani 1, I-40127 Bologna, Italy\label{Bol} \and
INAF - Osservatorio Astronomico di Roma, via Frascati 33, 00040, Monteporzio, Italy\label{Rom} \and
College of General Education, Osaka Sangyo University, 3-1-1, Nakagaito, Daito, Osaka 574-8530, Japan\label{osaka} \and
Observatoire de Genève, Université de Genève, 51 Ch. des Maillettes, 1290 Versoix, Switzerland\label{geneva} \and
CNRS, IRAP, 14 Avenue E. Belin, 31400 Toulouse, France\label{tolo} \and
Department of Astronomy, University of Massachusetts, 710 North Pleasant Street, Amherst, MA 01003, USA\label{UMASS} \and
Department of Physics and Astronomy, University of California, Riverside CA 92521, USA\label{river} \and
Dipartimento di Fisica e Astronomia, Universita' degli Studi di Bologna, Viale Berti-Pichat 6/2, 40127 Bologna, Italy\label{unibo}
}

   \date{Received -; accepted -}


  \abstract
  {}
   {We aim to investigate the effect of the escaping ionizing radiation
    on the color selection of high-redshift galaxies and identify candidate 
    Lyman-continuum (LyC) emitters.} 
   {We used the intergalactic medium (IGM) prescription of Inoue et al.~(2014) and galaxy synthesis 
    models of Bruzual~\&~Charlot~(2003) to properly treat the
    ultraviolet stellar emission and the stochasticity of the 
    intergalactic transmission and mean free path in the ionizing regime.
    Color tracks were computed by turning the escape fraction {\em fesc} of ionizing radiation on or off.}
   { At variance with recent studies,
     a careful treatment of IGM transmission leads to no significant effects on the high-redshift
     broad-band color selection, even adopting the most extreme
     ionizing emission model (with an age of 1Myr, zero dust, and metallicity $Z/Z_{\odot}$=0.02).
     The decreasing mean free path of ionizing photons with increasing redshift further diminishes 
     the contribution of the LyC to broad-band colors.
     We demonstrate that prominent LyC sources can be selected under suitable conditions by 
     calculating the probability of a null escaping ionizing radiation. This was performed by running 
     {\it ad hoc} Monte Carlo simulations anchored to the observed
     photometry, exploring the stochasticity of the IGM, and comparing 
     the simulated and observed colors that encompass the Lyman edge.
     The method was applied to a sample of galaxies extracted from the GOODS-S field.
     A  known LyC source at $z=3.795$ was successfully recovered as a LyC-emitter candidate, and 
     another convincing candidate at $z=3.212$ is reported. A detailed analysis of the 
     two sources (including their variability and morphology) suggests a possible mixture
     of stellar and non-stellar (AGN) contribution in the ultraviolet. 
   }
   { The classical broad-band color selection of $2.5<z<4.5$ galaxies does not prevent the inclusion 
     of LyC emitters in the selected samples.
     High {\em fesc} in relatively bright galaxies ($L > 0.1L^{\star}$) could be favored by the presence of a
     faint AGN that is not easily detected at any wavelength.
     A hybrid stellar and non-stellar (AGN) ionizing emission could coexist in these systems 
     and explain the tensions found among the UV excess and the stellar population synthesis models 
     reported in literature.} 


    \titlerunning{Focusing on Lyman continuum radiation}
   \authorrunning{Vanzella et al.}

   \maketitle

%

\section{Introduction}

In the past two decades, photometric redshift and color selection techniques 
have been widely used to identify high-redshift galaxies up to
the reionization epoch, $z>6$
(\citealt{ste99}; \citealt{bu04}; \citealt{giava04}; \citealt{vanz09}; \citealt{cas10}; 
\citealt{ag11}; \citealt{mclure13}; \citealt{fink12}; \citealt{bo14}).
The efficiency of the selection and the volume probed depend on the photometric
system, the depth of the survey \citep{duncan14},  the contribution of
nebular emission (e.g., \citealt{SB10}; \citealt{I11a}) 
and the intrinsic star-formation history
and duty cycle over which a galaxy becomes UV-bright and detectable in the ultraviolet
(\citealt{wlo14}). 
Potential contaminants include stellar sources, time-variable events 
like supernovae, and spurious sources (e.g., \citealt{bo14}). Interveneing lower redshift objects that are not spatially resolved (or not de-blended) can also alter the observed magnitudes.
Therefore, a well-defined cosmological volume (i.e., redshift interval)
does not have a well-defined corresponding region in color space.

The global shape of the spectral energy distribution and the 
discontinuities in the emerging electromagnetic spectrum
have been used to estimate the redshift probability distribution functions 
(i.e., photometric redshifts).
Typical observed breaks in the ultraviolet-optical regime are the Lyman edge at 912\AA, 
produced by the photoelectric absorption of neutral hydrogen in the stars, the 
interstellar medium and the intergalactic medium, the $Ly\alpha$-break (1215.7\AA)
caused by the redshift-dependent intergalactic transmission blueward
of the $Ly\alpha$ transition,
and the Balmer break ($\lambda \simeq 4000$\AA), a proxy of the age of the stellar populations.
The contribution of the nebular line emission has also been demonstrated
to be significant at high redshift by mimicking, attenuating, or boosting 
the expected underlying breaks (e.g., Balmer and/or $Ly\alpha$-breaks,
e.g., \citealt{SB09},2010; \citealt{DB14}; \citealt{vanz09})
or generating evident discontinuities (jumps) in the photometry, as measured in extreme
emission-line galaxies (e.g., \citealt{vdw11}; \citealt{amorin11}; \citealt{maseda14})
or in sources in which the signature of the $H\alpha$ emission 
line or the [O\,{\sc iii}]$\lambda 4959,5007+H\beta$ emission structure have been
detected in the Spitzer/IRAC channels  at redshift $z>3.8$ and $z>6.5$, respectively 
(\citealt{shim11}; \citealt{smit13}).
Broad absorption features like the 2175\AA~bump are also expected to introduce dips in 
the photometry if a Milky Way-type extinction curve is present at high redshift 
(e.g., \citealt{capak11}; \citealt{kimm13}).

Of these discontinuities, the Lyman edge (912\AA) in star-forming galaxies
is still subject of investigation. Currently, a solid and direct measure of the stellar Lyman
break at high redshift ($z>2.5$) has not been reported. 
This is relevant for characterizing sources responsible for hydrogen 
reionization and is still an active line of research at any redshift. 
A statistically significant sample of sources with a solid Lyman-continuum (LyC) 
detection and a comprehensive view on how it is spatially distributed relative to 
non-ionizing light are still lacking 
(e.g., \citealt{vanz12}; \citealt{nestor13}; \citealt{iwata09}; \citealt{inoue11b}; \citealt{siana10}). 
Observationally, this study
is further complicated by the opacity of the intergalactic medium (IGM), which
becomes significant with increasing redshift.
On the one hand, this opacity, coupled with the internal absorption in the galaxy,
allows us to select high-z galaxies through the Lyman-break technique mentioned above, 
on the other hand, it could prevent us from selecting sources with prominent
escaping ionizing flux. Recently, \citet{cooke14} concluded that a 
significant fraction of galaxies is missed in color-color selection diagrams
if the ionizing radiation escapes.
In practice, at redshift $2<z<5$ the required drop in the
Lyman-break selection method could intrinsically introduce a bias and
exclude LyC emitters from the selection.

We have three main aims in the present work: {\bf (1)}
we address the above possible bias in the high-z selection by investigating 
the effect of the intergalactic transmission on the photometric selection
of high-redshift star-forming galaxies when a leakage of ionizing radiation is present.
To maximize the effect, we assume that all ionizing photons escape along the line of sight ({\em fesc}=1). 
The nebular line emission and photometric scatter also perturb colors and alter the positions
in the color-color plane and have been extensively explored and quantified in the past
(e.g., \citealt{duncan14}; \citealt{SB10}; \citealt{dahlen10}). 
These effects are also discussed.
{\bf (2)} We present a method for selecting LyC emitters by focusing on the
detailed analysis of ultraviolet colors and deriving the probability of a 
null escape fraction of ionizing photons. This method  allows us to identify 
potential LyC candidates without the use of a dedicated survey.
It is presented and applied to a test sample of spectroscopically confirmed galaxies. 
{\bf (3)} We test this selection method on a sample of spectroscopically confirmed
galaxies for which additional photometric and spectroscopic observations
presented here allow us to verify the results of our selection technique.

In Sect.~2 the methodology is presented, and the results of the color-color selection
techniques are discussed in Sect.~3. The method is applied to a sample 
of galaxies lying in the GOODS-S field, and the results are reported in Sect.~4.
In Sect.~5 we discuss and summarize the results.

Errors are quoted at the $1\sigma$ confidence level unless otherwise stated.
Magnitudes are given in the AB system (AB~$\equiv 31.4 - 2.5\log\langle f_\nu / \mathrm{nJy} \rangle$).
We assume a cosmology with $\Omega_{\rm tot}, \Omega_M, \Omega_\Lambda = 1.0, 0.3,
0.7$ and $H_0 = 70$~km~s$^{-1}$~Mpc$^{-1}$.

\section{Methodology}

We derived the synthetic colors of high-z galaxies by adopting a Monte Carlo prescription 
of the IGM transmission and including spectral templates to properly follow the ultraviolet 
stellar light emerging from galaxies (a schematic view is shown in Fig.~\ref{method}). 
In particular, we included two effects:
(1) the treatment of the stochasticity of the IGM transmission both
in the ionizing and non-ionizing regimes, in particular, the redshift evolution of 
the mean free path of ionizing photons is properly considered; 
(2) we used galaxy templates that cover the
Lyman break and the ionizing emission, such that the shape of the Lyman continuum 
($\lambda<912$\AA) is suitably represented and is self-consistent with the whole
ultraviolet emission.  We did not include dust attenuation for two
reasons: first, although the dust extinction law would
proceed monotonically in the ionizing regime and therefore we expect more 
attenuation in the LyC than at 1500\AA~(e.g., \citet{siana07}, see their Figure~4), we prefer 
to avoid this extrapolation and assumed for now no dust effect on color-color selection.
Second, when we exclude dust, we favor a stronger 
effect of ionizing radiation on colors, therefore our approach is conservative.

\subsection{Intergalactic medium perscription}
\citet{II08} (II08 hereafter) calculated the transmission of the IGM by 
performing dedicated Monte Carlo (MC) simulations.
These simulations have been used to correct for intergalactic attenuation 
and estimate the escape fraction of ionizing radiation 
(e.g., \citealt{iwata09}; \citealt{vanz10b}, 2012). 
The MC simulation of II08 have now been updated based on the new 
empirical function of intergalactic absorbers presented in 
\citet{I14} (I14, hereafter).
We briefly recall the main steps. The simulations are based on an 
empirical distribution function of intergalactic absorbers that simultaneously
reproduces the observational statistics of the Lyman alpha forest (LAF), 
Lyman limit systems (LLSs), and damped Lyman-alpha systems (DLAs). From this assumed distribution function, a large number 
of absorbers were generated 
along many lines of sight. The probability of encountering an absorber was assumed 
to follow a Poisson distribution, and for each one the
column density and Doppler parameter were extracted randomly
from their (empirical) probability distribution functions. 

The simulations are therefore anchored to the observed distribution 
of the number density of absorbers at various column densities
(e.g., \citealt{pro09}, 2010, 2014; \citealt{songaila10}; \citealt{omeara13}), and verified to be 
consistent with the mean $Ly\alpha$ transmission
 (\citealt{becker13}; \citealt{FG08}) and with the mean free path of ionizing photons 
(\citealt{worseck14}) in the redshift interval 1-6 (see I14 for more details).
We also verified that the mean and scatter at $\lambda>912$\AA~was in line with that observed 
in QSOs studies (e.g., \citealt{FG08}; \citealt{rollinde13}).
The distributions of the median transmissions with the central 68\% intervals 
computed in the rest-frame ranges $880-910$\AA~and $1070-1170$\AA~are 
shown in Fig.~\ref{IGM}, in which the asymmetric distribution 
is evident in both cases with a tail toward low transmissions due to the presence of DLA and LL systems 
(see I14 and II08 Fig. 7 for details). As discussed in II08,  the absorption of the LyC is mainly
caused by LLSs and DLAs with $N_{HI}>10^{17}cm^{-2}$. This implies 
that LyC absorption is very stochastic because it is related to the
probability of intercepting dense absorbers (e.g., LLSs).
The simulations (which are based on the new absorber function by I14) 
used in this work consist of 
10000 random lines of sight at a given redshift, in the range $0-7$ with $dz=0.1$, each transmission 
spans the wavelength (rest-frame) interval 600-1300\AA~with $d\lambda$=1\AA~(an example
of a single l.o.s. is shown Fig.~\ref{method}).
Redward of the $Ly\alpha$ transition (1216\AA), the IGM transmission is assumed to be 1.0.
These simulations are therefore ideal to account properly for the effect
of the IGM stochasticity on color selection.

\subsection{Galaxy template spectra}

The ultraviolet rest-frame spectra of star-forming galaxies are well reproduced with 
spectral synthesis models (e.g., \citealt{ilbert09}; \citealt{heckman11}; \citealt{leit14}).
We adopted galaxy template spectra of \citet{BC03}(BC03, hereafter), 
which are widely used in the literature and  clearly describe the typical ultraviolet and ionizing
emission arising from star-forming galaxies.
At the level of accuracy required here, more sophisticated models of stellar evolution 
do not introduce significant differences in the results (e.g., \citealt{levesque12}).
We must rely on theoretical models because of the lack of observed spectra 
of confirmed galaxy with Lyman-continuum leakage below the Lyman limit.

\begin{figure}
\centering
\includegraphics[width=9.0cm, angle=0]{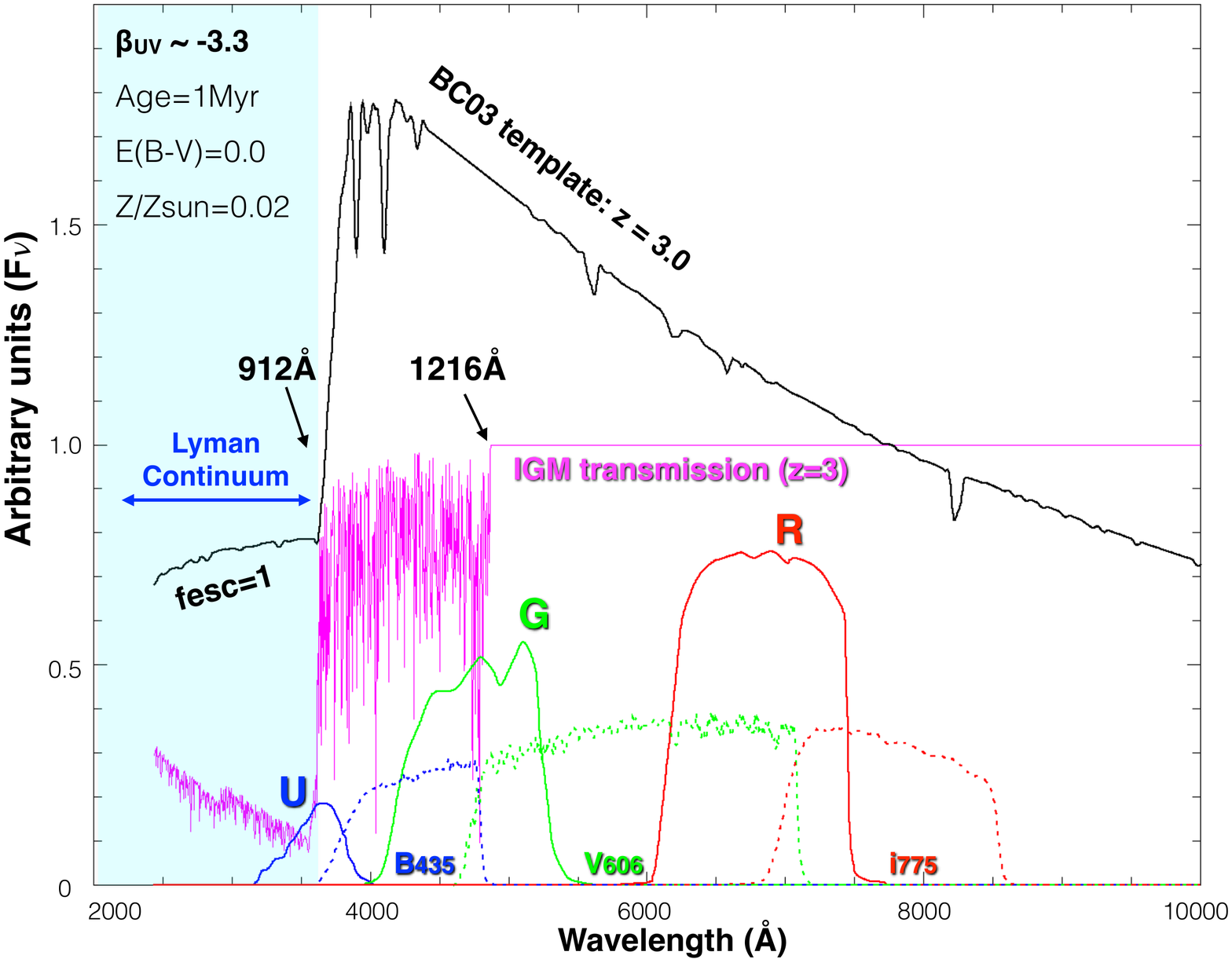}
\caption{Examples of the galaxy template ($F_{\nu}$, black line) 
placed at redshift $z=3.0$, intergalactic transmission (one l.o.s. out of 10000) 
and the filters used in this work. The Lyman continuum is indicated (shaded region), 
and in this example it affects half of the $U_{n}$-band. The BC template shown is the
most extreme used in this work and maximizes the ionizing emission 
(it corresponds to an ultraviolet slope of $\beta=-3.3$). }
\label{method}
\end{figure}

\begin{figure}
\centering
\includegraphics[width=9.0cm, angle=0]{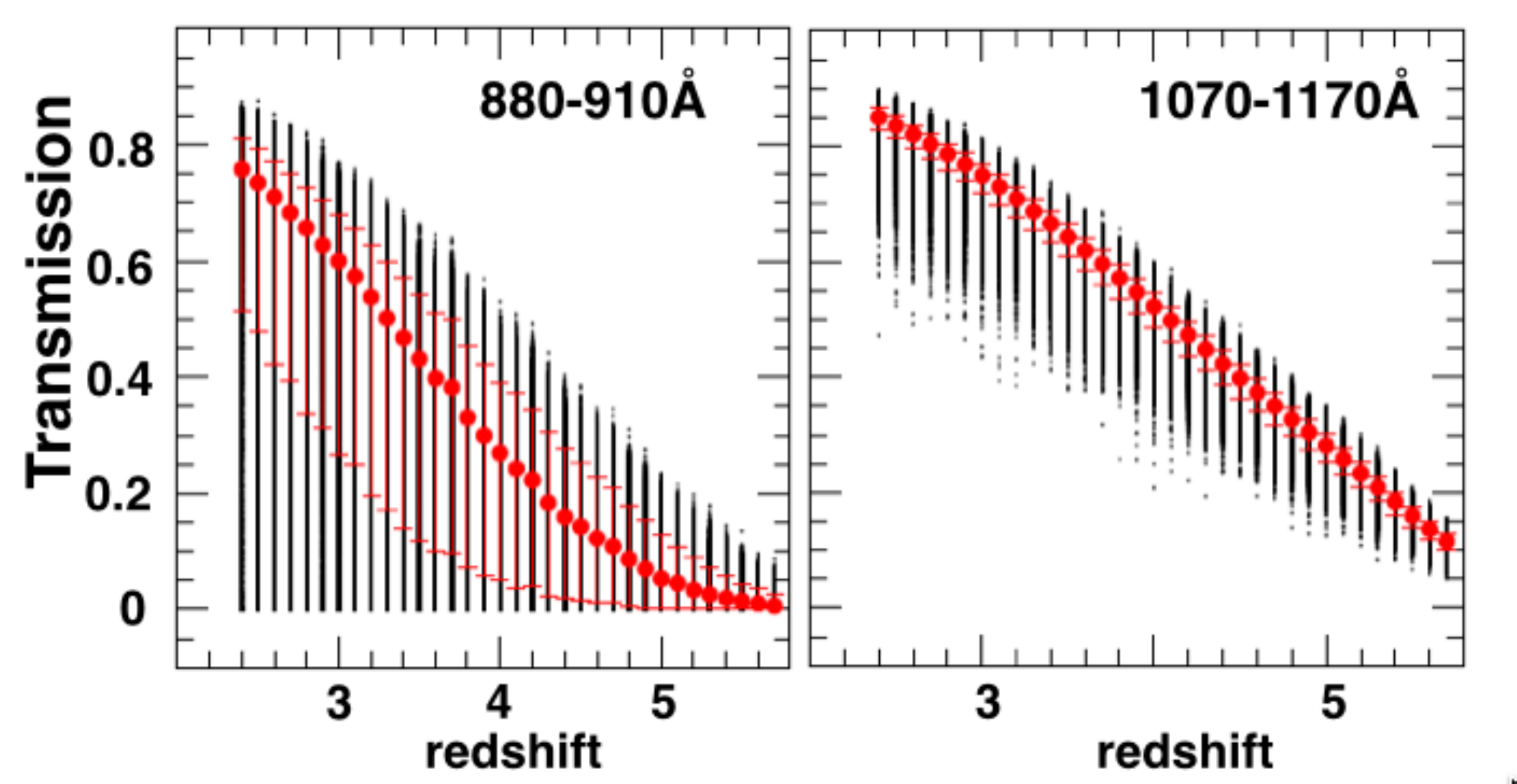}
\caption{IGM transmissions
as a function of redshift calculated in the wavelength intervals 880-910\AA~(left) and
1070-1170\AA~(right). At each redshift, 10000 line of sights have been computed
(black points), and the medians with the central 68\% intervals are
also shown (red filled circles with errorbars).}
\label{IGM}
\end{figure}

The shape and intensity of the ionizing radiation is linked to the nature of the stellar populations 
involved, such that for standard stellar populations the luminosity density ratio
$L(1500\AA)/L(\lambda)$ increases as wavelength decreases 
($\lambda < 912$\AA, e.g., \citealt{inoue05}; \citealt{siana07}).
An extreme case is shown in Fig.~\ref{method}, in which a galaxy
template placed at $z=3.0$ with zero dust extinction, E(B-V)=0, low metallicity $Z/Z_{\odot}=0.02$
and young, 1Myr old, is superimposed to the photometric filters used in this work. 
These parameters correspond to an ultraviolet slope of $\beta_{UV} \simeq -3.3$,
which is quite extreme and very rarely observed  (e.g., \citealt{vanz14a}).
The ultraviolet density ratio for this template is $F_{\nu}(1500)/F_{\nu}(900)=1.7$,
extremely blue if compared with typical star-forming galaxies at the same redshift
($F_{\nu}(1500)/F_{\nu}(900) \simeq 6$, \citet{siana10}).
However, we used this extreme case to maximize the effect of the ionizing emission 
on colors when an {\em fesc}=1 is assumed.
Even more extreme ionizing emission can be realized as in the Lyman-bump scenario 
described by \citet{inoue10}. Here we limit our analysis to the standard BC models.
In the same figure a random intergalactic transmission 
(one out of 10000 at $z=3$) is also shown (magenta solid line). In this case 
a relatively high IGM transparency is present down to $\simeq 600$\AA~rest-frame. 
A galaxy template with a redder and more typical UV slope  ($\beta_{UV} \simeq -2.1$) 
has also been adopted (E(B-V)=0, $Z/Z_{\odot}=0.2$, age=1Gyr).

%


%
\begin{figure*}
\centering
\includegraphics[width=15.5 cm, angle=0]{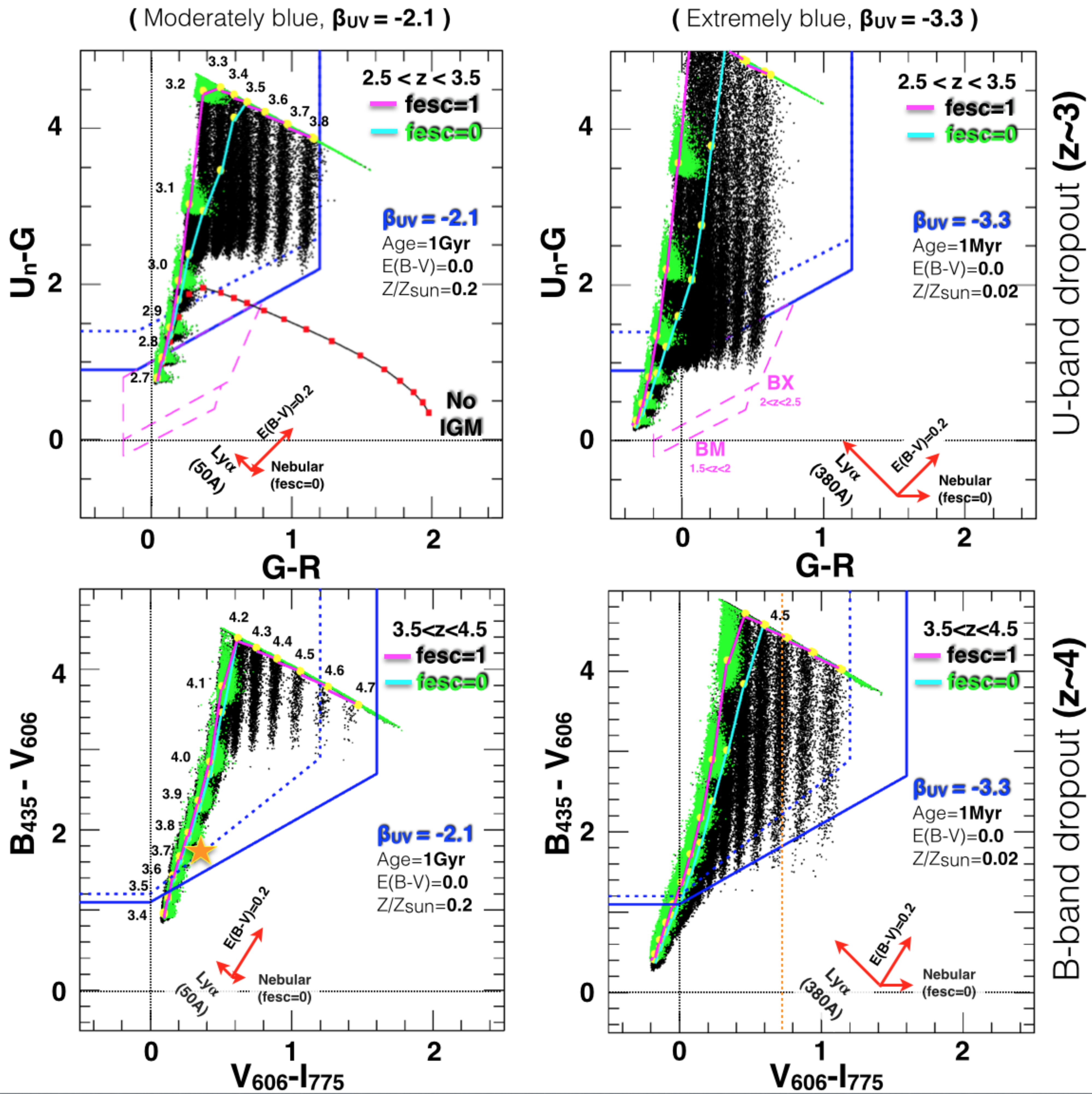}
\caption{{\bf Top panels:} The $U_{n}GR$ color planes with the typical selection regions for
galaxies (solid blue line) in the redshift range $2.5<z<3.5$ are shown for moderately blue (left) and extremely blue (right) galaxy templates. The effect of dust extinction and
$Ly\alpha$ emission is shown with red arrows. 
The green clouds and black stripes correspond to
{\em fesc}=0 and {\em fesc}=1, respectively, and are computed with redshift steps of 0.1 indicated at the 
left side. Magenta and cyan solid lines connect the barycenter of each cloud/stripe and represent the 
average color tracks (see text).  In the top-left panel the (solid) black color track with filled (red) squares denotes 
the case in which the IGM is absent from the ionizing part of the spectrum, i.e., the transmission is equal to 1.
The rightmost filled square symbol represents $z=4.5$; squares have been plotted with $dz=0.1$.
The color track resembles the behavior of the flat models of Cooke et al. (2014) (see Sect. 3.1).
Magenta dashed lines underline the BM and BX selection regions.
   {\bf Bottom panels:} The same as top panels, but in this case, the $BVI$ ($B_{435}, V_{606}, i_{775}$)  
color selections of galaxies 
 expected to lie in the redshift range $3.4<z<4.5$ are shown (solid blue lines). In the bottom-left panel the star indicates
the position of the Lyman-continuum emitters {\em Ion1} at $z=3.795$ (see text). 
In the bottom right panel the vertical orange dotted 
line highlights the bending of the black stripe at $z=4.5$ when ionizing
emission also affects the $V_{606}$-band, visible as a slightly bluer $V_{606}-i_{775}$ color for the higher IGM
transmission cases (see text).
The effect of the LyC leakage on colors decreases with increasing redshift (passing from $U$-band to $B$-band dropout
selection methods) as a result of redshift evolution of the mean free path of ionizing photons.
In all quadrants the triplets of red arrows mark the direction and magnitude of the color modification 
as a result of the $Ly\alpha$ emission, nebular continuum (affecting only the {\em fesc}=0 case, green points), 
and the dust attenuation, E(B-V).}
\label{colors}
\end{figure*}

\section{Results of the color selection}
$U_n-G$, $G-\cal R$ (\citealt{steidel03}), $B_{435}-V_{606}$, and $V_{606}-i_{775}$ (HST/ACS) colors were 
calculated at each redshift ($2.7-4.8$, $dz=0.1$) by convolving 
10000 IGM l.o.s. with filters and BC galaxy templates. The escape fraction of ionizing 
radiation has been turned off or on by imposing zero signal blueward of the Lyman 
edge ({\em fesc}=0) or by retaining the shape of the original BC template ({\em fesc}=1). 
The emerging 
ionizing  flux is consistent with the rest of the ultraviolet emission,
meaning that it is not artificially added.

Figure~\ref{colors} shows the results in the $U_{n}GR$ ($U_{n}-G$ vs. $G-R$) and $BVI$ 
($B_{435} -V_{606}$ vs. $V_{606}-i_{775}$) 
color-color planes typically used to select galaxies in the redshift range
 $2.5<z<3.5$ ($U_{n}$-band dropouts, \citealt{steidel03}) 
and $3.4<z<4.5$ ($B_{435}$-band dropouts, \citealt{giava04}). 
To set limits in the color diagrams, a one-sigma limiting magnitude of 30 (in all bands) was assumed
and the template spectrum was normalized to $z_{850}=25.0$. 
The saturation in the upper part of the color-color diagrams visible as a decreasing linear envelope is due to
this assumption. In this exercise we did not include photometric scatter 
(but see the discussion below).
While the dust absorbs the UV radiation,
both metallicity and age anticorrelate with the production of ionizing radiation, such that
the younger the stellar population and/or the lower the metallicity, the higher the
production of ionizing photons (e.g., \citealt{NO14}).

Two BC templates were adopted and are shown in Fig.~\ref{colors}: 
the extreme case already mentioned above ($\beta=-3.3$) aimed to
maximize the effect of LyC on colors, and another more typical
case (even though it is still relatively blue, $\beta=-2.1$).
We indicate in the same figure with an arrow (direction and magnitude) 
the effect that  $E(B-V)=0.2$ would have on colors by monotonically
extrapolating 
the law of \citet{c00} \citep[see also,][]{siana07}. 
The green and black points shown in Fig.~\ref{colors} were calculated adopting 
{\em fesc}=0 and 1, and are named green clouds or black stripes in the following.

{\bf [{\em fesc=0}]: }
Each green cloud shown in Fig.~\ref{colors} is made of 10000 points (that is, 10000 l.o.s.), and their
behavior represents the color regions with {\em fesc}=0 at various redshifts. 
The redshift intervals explored in the $U_{n}GR$  and $BVI$ diagrams are $2.7<z<4.1$ and $3.4<z<4.8$ (with $dz=0.1$), 
respectively. With increasing redshift, the Lyman edge and the null flux 
at $\lambda<912$\AA~enters the $U_{n}$-band, producing a rapid drop ($B_{435}$-band in the $BVI$ case). 
The spread in the horizontal and vertical directions reflects the IGM scatter in the non -ionizing domain, $\lambda>912$\AA. In the horizontal axis 
it only affects the G-band (V-band in the $BVI$ case), while in the vertical axis both $U_{n}$ and $G$ 
bands are affected. This scatter is in line with that observed in QSOs studies
(e.g., \citealt{FG08}; \citealt{rollinde13}; I14).
Focusing on the $U_{n}GR$ plane, the clear ($U_{n}-G$) color cut in the bottom envelope of the
green clouds is therefore due to a sharp
break at 912\AA~({\em fesc}=0), and the scatter in the (G-R) reflects
the scatter of the IGM at ($\lambda > 912$\AA).
The barycenters of the green clouds (connected with the magenta line in Fig.~\ref{colors}) resemble the typical 
color tracks reported in the literature, according to the adopted galaxy template and dust attenuation.

{\bf [{\em fesc=1}]: }
More relevant in this work is the case with {\em fesc}=1. The stochasticity of the IGM transmission in the LyC 
region and its effects on colors is appreciable. For example, the tail of transmissions toward high values 
(transparent l.o.s.) is clearly visible in the $U_{n}GR$ diagram as a percolation of points toward relatively 
blue ($U_{n}-G$) values (vertical black stripes). This is more evident in the extreme case ($\beta_{UV}=-3.3$), 
in which the ionizing emission is maximized in the set of models we used.
 We note that the median color tracks for {\em fesc=0/1} can differ
in the $U_{n}-G$ color in the $U_{n}GR$ plane by a negligible
quantity up to one magnitude in the $\beta=-2.1$ case. 
It is even higher for the most extreme template ($\beta=-3.3$)
because the ionizing flux is higher. 
This difference decreases as redshift increases 
(because of the decreasing mean free path of ionizing photons with redshift),
as shown in the $BVi$ $z\simeq4$ plane (see Fig.~\ref{colors}). 
However, and most importantly, the escaping ionizing radiation does not
significantly move the points outside the selection regions.
In particular in the $\beta=-2.1$ case,
there are no solutions outside the selection region as a result
of the ionizing emission. The solutions
escaping the region in the low-redshift tail, $z<2.8$ ($z<3.5$ in the BVI case) are due to 
the intrinsic blue slope, regardless of {\em fesc}. In the more extreme case, $\beta=-3.3$, 
only a marginal fraction leaves the selection box ($U_{n}GR$) at $z>3.7$, a redshift range that
is also covered by the other higher-z $BVI$ color selection,
however. In the $U_{n}GR$ and $BVi$ diagrams the
fraction is always $<10\%$ in the $\beta=-3.3$ BC template. However, the selection boxes were designed to capture star-forming galaxies with 
typical properties, therefore it is not surprising that the $\beta=-3.3$ case does not exactly match the redshift range, especially the lower redshift edge of the selection volume.
We recall that the two templates adopted here are dust-free and tend to maximize the contribution
of ionizing emission to the colors.

The effect of an escaping ionizing radiation ({\em fesc}=1) decreases as redshift increases, mirroring the
redshift evolution of the mean free path of ionizing photons. For this reason, the same black
stripes in the $BVI$ plane (redshift $3.4-4.5$) are less prominent than in the lower z $U_{n}GR$ plane.
If the m.f.p. and its redshift evolution are neglected from the calculation, the dropout signature
is strongly attenuated and color tracks escape from the selection region (see next section).

A second-order effect of {\em fesc}=1 is a small bending of the stripes in the case 
of high IGM transmissions if the Lyman edge affects both colors in the given 
color-color diagram ($U_{n}GR$ or $BVi$). 
An example is indicated with a vertical orange dashed line (Fig.~\ref{colors}, bottom right panel) 
superimposed to the black stripe at $z=4.5$. A bending in the bottom envelope of the stripe 
at ($B_{435}-V_{606}$)$~\simeq2.0$ and ($V_{606}-i_{775}$)$~\simeq0.7$ is appreciable. 
This is because the LyC 
also boosts the $V_{606}$-band, producing slightly bluer values
($V_{606}-i_{775}$). 
However, this is a second-order effect that can be easily erased by photometric noise.

\subsection{Marginal effect of the escaping LyC on the broad-band color selection}
As shown above, the effect of an escaping ionizing radiation on the color selection is minimal.
Conversely, \citet{cooke14} found a significant contribution and calculated color tracks that
lie outside the selection regions when the escaping ionizing radiation is present.
They performed spectrophotometry on composite spectra based on $z\sim3$ LBGs 
(e.g., \citealt{shapley03}) and applied various levels of escaping ionizing flux 
down to a $\lambda_{MIN}$, and assuming various $f1500/f_{LyC}$ ratios.
Two models have been adopted: 1) the case in which $\lambda_{MIN}$ is not
specified (called flat-model) and 2) the case in which
$\lambda_{MIN}=800$\AA~rest-frame (called step-model). While the adopted 
non-ionizing spectrum is (by construction) close to the average of the 
observed population, the assumptions on the Lyman continuum are too
simplistic and lead to unrealistic color tracks, especially in the flat-model. 
The flat-model neglects the m.f.p. and its redshift evolution, 
therefore the dropout signature is significantly reduced. The color tracks plotted 
over the $U_{n}GR$ plane in Cooke et al. (2014) (see, e.g., their Fig. 3 and 4) show
colors outside the selection region, reaching the extreme case at $z=4.5$ in which 
the $U_{n}-G$ is close to zero,  that is, with increasing redshift, 
the LyC enters the $G$-band and still survives in the $U_{n}$ band. 
At $z=4.5$ the $U_{n}$-band probes
$\lambda<730$\AA~rest-frame, while the m.f.p. at this redshift is 22Mpc (\citealt{worseck14}),
and corresponds to 880\AA~rest-frame, that is, on average, an ionizing photon 
survives on a much shorter path than is allowed in the flat-model.
The top-left panel of Fig.~\ref{colors} shows the color track of the BC template 
with $\beta = -2.1$, {\em fesc}=1  and assuming 100\% IGM
transmission blueward of the Lyman edge ($\lambda<912$\AA, solid black line and red squares).
The behavior of the flat-model is well reproduced 
(upturned shape) with the dropout in the $U_{n}$-band that never occurs 
up to $z=4.5$ (as in Cooke et al.). As discussed above, neglecting the m.f.p. of ionizing 
radiation in the intergalactic medium leads to a strong attenuation of the observed drop.
The comparison between the cyan and red tracks in the same figure
shows the strong effect that is achieved when the IGM is neglected in the ionizing domain.
The step-model is more realistic than the flat-model and produces color tracks
typical of what has been reported in literature (e.g., Steidel et al.), although, again, it does
not include the redshift evolution of the m.f.p. of ionizing photons.
The m.f.p. at $z=3.0/3.5/4.0$ is $\simeq 100/50/33$Mpc proper (\citealt{fuma13}; \citealt{worseck14}), 
which corresponds to $\lambda_{MIN} \sim 820/858/870$\AA~rest-frame. Therefore the assumption of 
$\lambda_{MIN}=800$\AA~is valid only for a narrow redshift range around $z \lesssim 3$
and tends to underestimate the growth-rate of the $U_{n}-G$ color with increasing redshift ($z>3$).

We here properly treat the aforementioned limitations, 
that is, the shape of the LyC flux (modeled with galaxy BC-templates) 
and the m.f.p. of ionizing photons (statistically described by the I08,I14 prescription) 
are both included in the MC simulations. Our results show that there is no significant
effect of ionizing leakage on the broad-band color selection. However, as discussed in
Sect.~4, a leakage of ionizing radiation is in principle identifiable under
suitable conditions. 

Another effect that can decrease the photometric breaks or change the UV slope
in galaxies and move points toward the outer part of the selection regions
is the (photometric) contamination by lower redshift superimposed sources,
which is not negligible in the case of deep and/or low spatial resolution surveys
(e.g., see \citet{vanz10a} for a statistical analysis).
Taken as a whole, it is worth noting that this type of contamination is different from
the effect produced by a pure escaping ionizing radiation. The latter is modulated
by the redshift evolution of the IGM opacity (m.f.p.) and therefore 
limited in its global contribution to the band. Conversely, the contribution of a 
lower redshift source mimicking  the LyC emission 
typically affects the entire band(s) blueward of the Lyman edge,
which significantly reduces the observed drop or causes the background higher-z galaxy 
to not drop at all, similarly to what was calculated in the flat-model of Cooke et al.

\subsection{Photometric noise, nebular continuum, and emission lines}

The effect of photometric noise on color selections has been extensively 
explored in the literature (e.g., \citealt{dahlen10}; \citealt{duncan14}).
The photometric scatter depends on the signal-to-noise ratio
(S/N) in the bands, which in turn is related to the depth of the survey and/or to the 
magnitude of the sources. In the present case - at fixed conditions - 
the {\em fesc}=0 produces fainter $U_{n}$-band (or $B_{435}$-band)
magnitudes than the {\em fesc}=1 (by definition), therefore
we expect that the {\em fesc}=1 case is slightly 
less scattered than the {\em fesc}=0.
It is beyond the scope of the present work to address this question or replicate
the analysis performed, for instance, by \citet{duncan14}. We only 
recall that depending on the adopted depth of the survey and magnitude 
of the sources, the photometric scatter can move points outside the selection
boxes, and vice versa. \citet{duncan14} found that most of the cases that escape 
the selection regions are photometric limits, meaning that they are compatible
with the high-z locus in the color space
(with virtually infinite photometric accuracy, they would be included
in the high-z regions).
Clearly, the high-z selection is futile above a certain level of noise.  

The $Ly\alpha$ line in emission can also affect the observed colors. 
Since the contribution to the broad-band is proportional to the observed equivalent 
width of the line, this effect is more significant with increasing redshift.
Given the redshift interval explored here ($2.5<z<4.5$), the contribution 
of the $Ly\alpha$ emission is relatively small and anyway favors the color-color 
selection criteria, that is, it tends to move points toward the selection region.
For the$U_{n}GR$ color plane 
the $Ly\alpha$ line always lies in the $G$-band and therefore any emission
moves points to the upper left part of the diagram. As an example, the
equivalent width of 65\AA~rest-frame
(corresponding to the highest theoretically expected value in the $\beta=-2.1$ 
case adopted here and assuming case B recombination)
introduces a differential magnitude $dm \simeq 0.15$ in the ($U_{n}-G$) and ($G-R$) colors, 
in the redder and bluer directions, respectively (the effect is shown as a vector in Fig.~\ref{colors}).
The same occurs for $BVI$ . Clearly, if the $Ly\alpha$ line enters the $U_{n}$-band 
(or $B_{435}$-band in the $BVI$ case), the effect is the opposite: the line decreases the
observed break. However, this is not the case here.
In the extreme case adopted here (with $\beta \simeq -3.3$), the effect of a $Ly\alpha$ line
with rest-frame equivalent width of 375\AA~is more significant, as expected. The nebular continuum
can also redden the UV slope (e.g., \citealt{cas14}). It is maximized in the {\em fesc}=0 case, in which the
ionizing radiation concurs to boost flux in the non-ionizing UV range (\citealt{sch02}; 
\citealt{inoue11b}). 
In the moderate case  ($\beta=-2.1$), the nebular continuum effect is negligible,
producing a $\Delta \beta = +0.07$, while in the $\beta=-3.3$ case a $\Delta \beta = +0.60$ is obtained, 
clearly in line with the extreme physical conditions 
(the effect on colors is reported with an arrow in Fig.~\ref{colors}).

\section{Photometric selection of LyC candidates}

Ionizing sources have been searched for by directly examining the 
Lyman continuum emission at $2<z<4.5$ with deep spectroscopy 
(\citealt{steidel01}; \citealt{giallongo02}; \citealt{shapley06}),
narrow-band imaging (e.g., \citealt{iwata09}; \citealt{nestor13}; \citealt{mostardi13}),
and deep intermediate and broad-band 
imaging (e.g., \citealt{ferna03}; \citealt{siana07},~2010; \citealt{vanz10b},~2012; 
\citealt{boutsia11}; Grazian et al. 2015). Deep slitless spectroscopy
has also provided deep limits at $z\sim1$ \citep{bridge10}.

As discussed in the previous section, the leakage of stellar ionizing 
radiation ({\em fesc}=1) convolved with the IGM does not compromise 
the color selection of high-z galaxies. Despite this, 
a statistically significant sample of LyC emitters ({\em fesc>0}) at high 
redshift is still lacking (e.g., see \citealt{vanz12}). 
One reason could be 
that a significant number of ionizing sources exist only in a much 
fainter and still unexplored luminosity domain (e.g., \citealt{wise14}; \citealt{kimm14}),
while relatively bright ($L>0.1L^{\star}$) ionizing sources are rare.
The contribution to the ionizing background therefore could be
negligible if we focus on the bright part of the luminosity function,
but identifying them is still interesting since the physical 
mechanisms that allow ionizing photons to escape are not fully 
understood and are probably linked to efficient feedback processes
(e.g., AGN).

In the following we describe a method for identifying LyC candidates
and apply this to a sample of spectroscopically confirmed
galaxies in the GOODS-S field. We
defer a 
systematic study on a larger spectroscopic redshift 
sample to a forthcoming work.

\subsection{Searching for anomalous UV colors as a signature of 
an escaping ionizing radiation}

As shown above, an optimal characterization of the stochasticity of the IGM 
and a suitable modeling of the ultraviolet emission of star-forming 
galaxies allows us to investigate the photometric contribution
of an escaping LyC radiation in more detail.
In particular, if the photometric accuracy is high enough, it is possible to
explore the position of the source in the color-color plane, focusing on the 
color(s) encompassing the Lyman edge, and determine whether it significantly departs from 
the {\em fesc}=0 assumption. 

To this aim, we built a set of templates for which we adopted different values of the ages, 
metallicities, and dust attenuation. 
Here the dust attenuation is inserted 
because we compare simulated with observed colors (discussed
in the next sections).
The following grid was adopted: 
$Z/Z_{\sun}$=1,0.2, and 0.02; E(B-V)=0.0,0.03,0.06,0.1,0.2,0.3,0.4,0.5,
and 0.6;
age=1Myr up to 2Gyr with steps of 0.05 in logarithmic scale.
Both constant and exponentially declining (with $\tau=0.6 Gyr$) star formation 
histories were assumed.
The Salpeter initial mass function and Calzetti  attenuation law 
extrapolated down to the LyC were adopted (e.g., \citealt{siana07}). 
The final sample uniformly covers the age, metallicity, 
and dust attenuation parameter space.

The basic idea is the following.
Given a galaxy with known spectroscopic redshift ({\it zspec})\footnote{We have shown 
that the photometric selection does not preclude the inclusion of LyC emitters.},
the UV spectral slope  ($\beta_{obs}$) with its one-sigma uncertainty ($\sigma_{\beta}$) is
measured by computing a multi-band fitting as described
in \citet{cas12}. The uncertainty on the estimated slope, $\sigma_{\beta}$, contains the 
photometric uncertainties.
The same procedure is performed on BC-templates to
derive the UV slope ($\beta_{model}$) at the same redshift and using the same bands. 
Subsequently, a subsample of galaxy templates with $\beta_{model}$ compatible
with the observed one is selected such that 
$\beta_{obs}-\sigma_{\beta}<\beta_{model}<\beta_{obs}+\sigma_{\beta}$.
To be conservative, we furthermore select in 
this sample the template that maximizes the ionizing output, by measuring 
the $L1500/L_{LyC}$ flux density ratio. This also means
that we select the template that maximizes the contribution of the non-ionizing
ultraviolet emission redward of the Lyman edge (e.g., $>912$\AA). This part still
affects the band even if {\em fesc}=0.
In this way, we build 
a galaxy template {\it anchored} to the observed UV slope and with the highest
possible (modeled) ultraviolet emission.
Then we convolve this template with 10000 IGM transmissions at the closest 
available redshift in the IGM redshift grid, $z_{IGM}$ (always $zspec-z_{IGM}<0.05$)
and derive magnitudes and colors, turning the escape fraction on and off, and normalizing 
the template to the observed $z_{850}$ magnitude.
In particular, we focus on the realization that assume {\em fesc}=0 and compare them with 
the observed photometry, with the aim to identify a possible excess of the flux 
in the ionizing domain.

The hypothesis is that we have no direct coverage just blueward of the Lyman
edge (912\AA), therefore we have no direct probe of the LyC region, 
as is the case in general for {\it non} -dedicated surveys. We wish to infer a signature of the 
ionizing radiation by focusing on the band in which the LyC is partly covered.
 While dedicated surveys based on narrow-band imaging and deep spectroscopy are more 
efficient in capturing the possible escaping ionizing flux just blueward of the Lyman limit
(although they are time-consuming 
and/or limited to narrow redshift intervals), the method described here complements them
and allows us  to explore possible LyC leakage in a continuous redshift range, exploiting existing
deep multifrequency surveys. The mean free path of ionizing photons is $\sim 100$Mpc 
physical at redshift 3 (\citealt{fuma13}; \citealt{worseck14}), which corresponds to $\sim$ 80\AA~rest-frame
blueward of the Lyman edge at the same redshift, that is,  $\sim$ 320\AA~in the observed-frame.
If compared to the typical width of broad-band filters, this is not negligible (but not dramatic
for the high-z selection functions, Sect.~3).

For example, the ($U$)$B_{435}$-band starts to include the LyC domain 
at $z>$($2.30$)$3.05,$ and at $z=$($3.0$)$3.8,$
half of the ($U$)$B_{435}$-band probes ionizing radiation. 
The distributions of the simulated colors and magnitudes derived assuming {\em fesc}=0 and 
containing the Lyman edge ($912$\AA) are then compared with the observations. 
In the case in which the $B_{435}$-band contains the Lyman limit, the 
probability that at the given observed color $(B-V)_{obs}$, UV slope $\beta$ and spectroscopic
redshift (with their uncertainties) is associated with an {\em fesc}=0 ({\em P(fesc=0)}) is calculated as follows:

\begin{equation}
P($fesc=0$)|_{[z_{spec},~\beta_{obs}\pm1\sigma_{\beta},~(B-V)_{obs}+1\sigma_{(B-V)} ]} = N_{[color]} / N_{[total]}
,\end{equation}

where $N_{[total]}=10000$ and $N_{[color]}$ is the number of realizations with {\em fesc}=0 
that reproduces the observed color, that is, $(B-V)_{simul} < (B-V)_{obs}+1\sigma_{(B-V)}$,
calculated adopting the BC galaxy template with the strongest LyC emission
of those that are compatible with the observed ultraviolet slope
$\beta_{obs}$ ($\pm 1\sigma_{\beta}$).
Therefore, given the observed 
quantities (color, redshift, and UV slope), a proper treatment of the
IGM transmission (I14) and UV emission (BC-models), the probability 
of a null {\em fesc} can be estimated from MC simulations. In 
particular, the LyC candidates are those with the lower conditional probability
 {\em P(fesc=0)}.

We applied this procedure to a sample of galaxies
   in the GOODS-S field with known spectroscopic redshift
   (\citealt{vanz08},~2009; \citealt{balestra10}). In this first
   test-run we focused on the $2.9<z<4$ redshift interval.
   The lower redshift limit was chosen such that  more than half of 
   the (ground-based) $U$-band probes the LyC. \footnote{Bluer bands like
   the ultraviolet channels provided by the HST/WFC3, e.g., F225W, F275W, and F336W,
   will allow us to investigate lower redshift regimes.} The upper limit is mainly 
   given by the short
   mean free path of ionizing photons at high redshift, $z>4$ (m.f.p. $<35$Mpc proper), 
   which makes the contribution to broad-band filters negligible on average.
   For each source the {\em P(fesc=0)} was calculated using the ($U-B$) 
   and (B-V) colors as described above.
   At $z>3.4$ the Lyman limit moves redward of the red edge of the $U$-band filter,
   which means that the $U$-band is a direct probe of the LyC.  
   These cases have previously been explored in detail in \citet{vanz10b}, where
   only one convincing LyC emitter (out of 102 LBGs) was identified
   and named {\it Ion1} (see below).
   Several other $U$-band detections at $z>3.4$ have been identified 
   in the same GOODS-S field,  but a careful
   investigation that exploited the optical and near-infrared high spatial
   resolution imaging (GOODS+CANDELS)  has revealed a probable contamination 
   by foreground sources that mimics  the LyC 
   (each of them analyzed in \citealt{vanz12}). 
   The fraction of contaminants is also compatible with that expected 
   from statistical arguments (see \citealt{vanz10a}).  

   In this application we furthermore selected sources with a $S/N > 5$ in all
   the ultraviolet bands and with secure spectroscopic redshift measurements.
   We obtained a sample of 32 galaxies in the range $2.9<z<4$, magnitude
   $24<z_{850}<25.5,$ and ultraviolet colors redward of the $Ly\alpha$ typical of LBG 
   populations (e.g., \citealt{vanz09}). The probabilities we derived ({\em P(fesc=0)})
   range between 0 and 100\%, with some cases of foreground
   contamination (discussed in \citealt{vanz12}) and known X-ray detected AGNs.
   In the following we focus on the sources with the lowest probability {\em P(fesc=0)}$<5\%$
   that are isolated, that is, those without evident contamination from nearby lower z objects,
   deferring a systematic application on a larger spectroscopic sample 
   and a detailed comparison of LyC emitter candidates
   and non-candidates to a future work. Here we aim to demonstrate the feasibility of the method.
   In this way, two out of 32 sources show {\em P(fesc=0)}$<5\%$:  one is the already known {\it Ion1} 
   source at $z=3.795$ (see next section and \citealt{vanz12}), the other is a new 
   intriguing candidate ionizer (see Sect.~4.3).
   These two objects are investigated in more detail in the following.

\subsection{LyC emitter: {\it Ion1}}

\begin{figure}
\centering
\includegraphics[width=8.5 cm, angle=0]{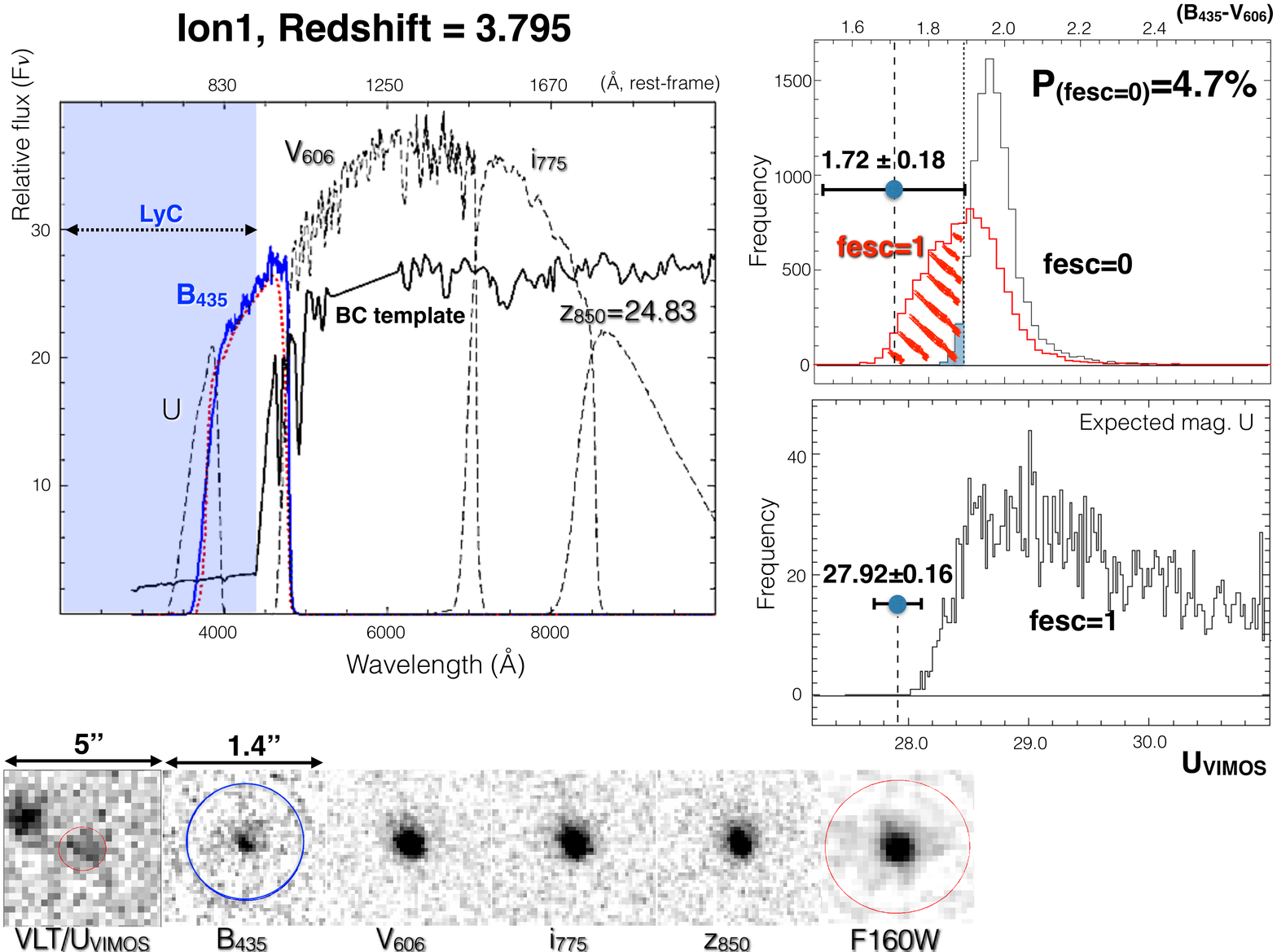} 
\caption{Summary of the {\it Ion1} source. 
In the top-left panel the HST/ACS, VLT/$U,$ and $B$-band filters are shown,
superimposed to a BC galaxy template placed at the same redshift of {\em Ion1}.
In particular, the solid blue and red dotted lines represent the 
HST/$B_{435}$ and VLT/$B$ band filters.
 The shaded transparent blue region marks the LyC region. More than half of the
 $B_{435}$ and VIMOS/B bands probe the ionizing domain, while the $U$-band probes
$\lambda<830$\AA~rest-frame. In the top right panel the simulated
distribution of the ($B_{435}-V_{606}$) colors are calculated assuming {\em fesc}=0/1 and compared
to the observed color of the source {\it Ion1}. The probability of a null {\it fesc} is very low.
Similarly, the simulated $U$-band magnitude reproduces
the observed one only marginally (middle right panel).
The bottom panel shows the thumbnails for {\it Ion1} . The blue circle
in the $B_{435}$-band has a diameter of $1^{\prime\prime}$.
}
\label{ion1}
\end{figure}

The procedure described above recovers {\it Ion1}  as a LyC emitter candidate (CANDELS ID 23836, \citealt{guo13}),
even excluding the direct information we have about the LyC detection (i.e., the $U$-band
detection\footnote{Here the $U$-band refers to the VIMOS $U$-band, \citet{nonino09}, which is an
intermediate-band filter, FWHM=350\AA~(see also \citealt{vanz10b}).}
at $\lambda < 830$\AA~rest-frame, Fig.~\ref{ion1}).

We recall that, currently, the global spectral energy distribution of this source is compatible
with a star-forming galaxy: there is no X-ray detection in the 4Ms Chandra observations,
and no emission lines (e.g., $Ly\alpha$, NV, CIV) have been measured in the spectrum (\citealt{vanz10b},~2012).
It is a compact source, but is marginally resolved in the ACS images ($\lambda \sim 1500$ rest-frame) 
with a $R_{eff} \simeq 250$ pc  (see Sect.~4.6).

At the redshift of {\it Ion1} more than half of the $B_{435}$ band probes the ionizing part of the
spectrum ($<912$\AA) (see top-left panel of Fig.~\ref{ion1}). The UV slope measured with the 
five-band fitting procedure gives $\beta_{obs}=-2.01 \pm 0.10$.  The observed color
is $(B_{435}-V_{606})=1.72$ with an error $\sigma_{(B-V)}=0.18,$ the $z_{850}$-band 
magnitude is $z_{850}=24.83 \pm 0.03$. 
We derived the probability {\em P(fesc=0)} by calculating the expected $(B_{435}-V_{606})_{simul}$ color 
including the I14 IGM prescription and adopting the galaxy (BC) model anchored to the observed 
UV slope, as discussed in the previous section.
The distribution of the expected color $(B_{435}-V_{606})_{simul}$ for the {\it Ion1} source
is shown in Fig.~\ref{ion1} and is compared to the observed color.
Interestingly, although this source was selected as a $B_{435}$-band dropout 
(indicated with a star in Fig.~\ref{colors}, bottom-left panel), the observed $(B_{435}-V_{606})$ 
color is reproduced only marginally if {\em fesc}=0, that is, {\em P(fesc=0)}$~\simeq 4.7\%$
(shaded region of the {\em fesc}=0 histogram shown in Fig.~\ref{ion1}, upper right panel).
With such a low probability it is selected as a LyC emitter candidate. 
Conversely, the color is better reproduced with the distribution that assumes {\em fesc}=1 
(see the red curve in the same figure). We recall that the adopted template is the most ionizing while
reproducing the observed UV slope. If we run the same procedure on all the templates 
that satisfy the condition $\beta_{obs}-\sigma_{\beta}<\beta_{model}<\beta_{obs}+\sigma_{\beta}$
 (239 BC templates)  regardless of any constraint on the ionizing emission,
$2.39 \times 10^{6}$ color estimates ($=N_{[total]}$) are derived. In this case, {\em P(fesc=0)} 
is 0.4\%. 

A similar analysis can be performed on the observed magnitudes that probe the LyC domain. 
As described above, the galaxy template was normalized to the observed $z_{850}$. 
While the expected magnitude in the $B_{435}$-band better matches the {\em fesc}=1 distribution
(although its error still prevents us from clearly distinguishing between {\em fesc}=1/0), 
the observed $U$-band magnitude of $27.92 \pm 0.16$  - that probes the $\lambda<830$\AA~rest-frame - 
is more difficult to reproduce {\em fesc}=1 and given the observed UV slope, even in the more 
transparent IGM l.o.s. and {\em fesc}=1, see Fig.~\ref{ion1} (the case with {\em fesc}=0 produces zero 
flux in the $U$-band by definition). The measured $U$-band flux can be better explained 
with  bluer and more ionizing BC templates (e.g., $\beta<-2.5$), which contradict the observed $\beta=-2.01 \pm 0.10$, however. This could suggest a non-stellar radiation that 
progressively dominates at shorter wavelengths 
(especially in the ionizing domain). In particular, if we adopt a simple underlying power-law 
template without any break at the Lyman edge (AGN-like), $F_{\lambda}=\lambda^{\beta}$ with $\beta=-2$,
we obtain even bluer ($B_{435}-V_{606}$) colors and brighter $U$-band 
magnitudes than observed. Obviously, pumping flux into the
ionizing regime allows us to recover the $U$-band magnitude as
well. 
We recall that this object is currently not detected in X-ray (the 4Ms Chandra observations 
provide a $1\sigma$ limit of $L_{X}<3.5\times10^{42}~erg~s^{-1}$)
or in the $24\mu m$ Spitzer/MIPS
observations \citep{vanz12},  and it is spatially resolved in the HST/ACS images
(see Sect.~4.6).
A possibility is that a mix of stellar and (faint) AGN  emission contributes to the ultraviolet spectrum.
More in general, it could be that relatively bright ionizing sources host a faint AGN that
favors the escaping ionizing radiation by decreasing the column density of neutral hydrogen gas
along the line of sight, causing {\em fesc} to be higher than zero, as in the {\em Ion1} case 
(i.e., $N_{HI}<10^{18} cm^{-2}$). This would imply that faint AGNs can still
provide the necessary feedback to increase the transparency toward certain view angles 
(e.g.,  \citealt{giallo12}). The ionizing photons produced by young stars in the compact central
region (revealed in the UV at $\lambda>912$\AA~as spatially resolved emission, see Sect.~4.6) 
could share a similar (or the same) physical path of non-stellar ionizing emission (AGN) and escape 
from the interstellar and the circum-galactic media.
In this scenario, the final escaping ionizing radiation would have a stellar and non-stellar
origin, with a certain balance among the two, possibly depending on the feedback 
mechanisms performing differently at different luminosity and
mass regimes.

Deeper data with HST/WFC3 UV channels would be needed  to examine the spatial distribution 
of the ionizing radiation (as a probe of a stellar ionizing emission), and/or deeper 
spectroscopic observations would be necessary to study the covering fraction of neutral 
gas through absorption line analysis  (e.g., \citealt{heckman11}; \citealt{borta14}).
\begin{figure}
\centering
\includegraphics[width=8.5 cm, angle=0]{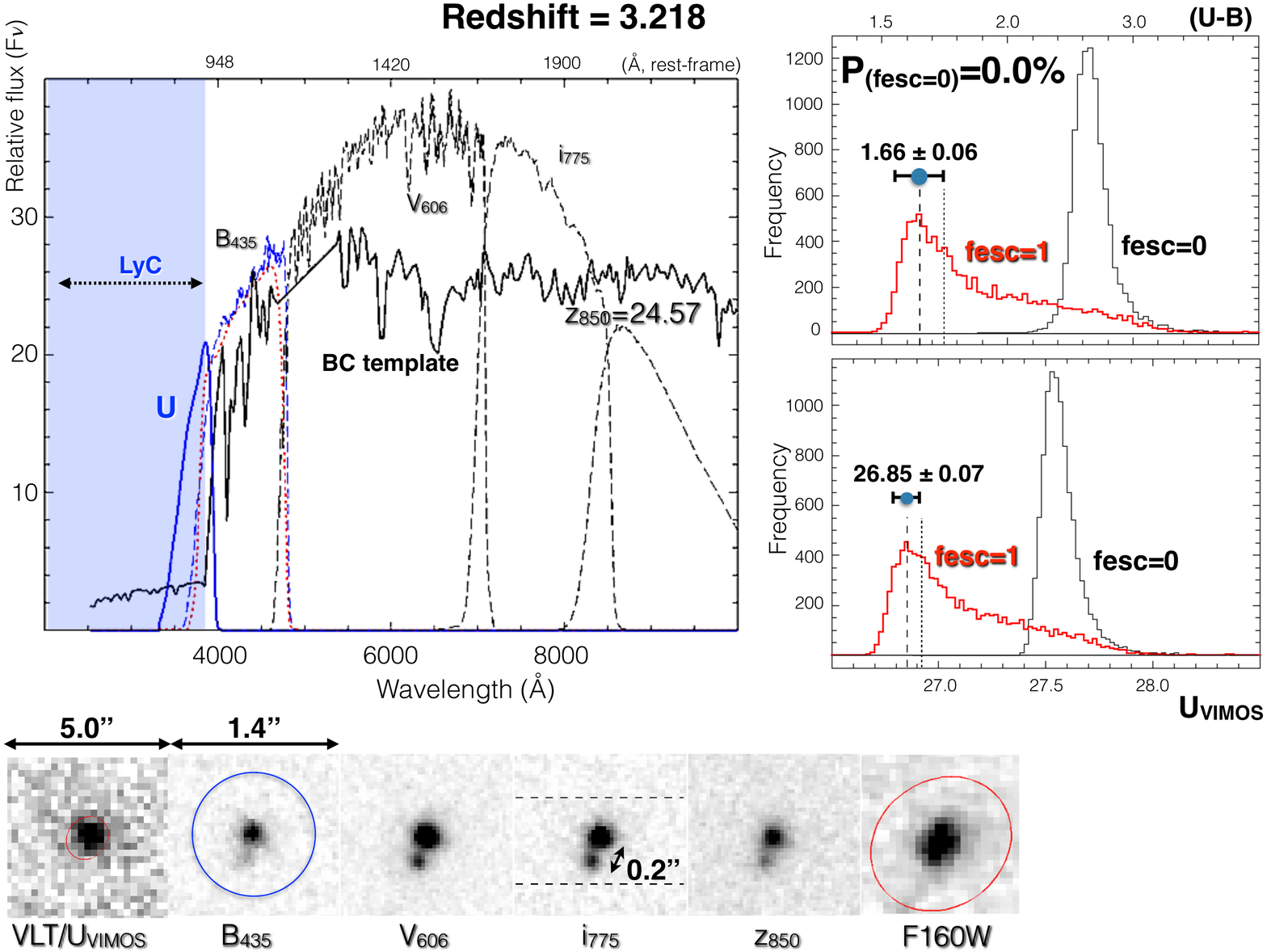}
\caption{Summary of the {\it Ion2} source, see description of Fig.~\ref{ion1}.
The slit orientation is shown in the $i_{775}$ image with the wavelength dispersion
along the vertical direction.}
\label{CS2}
\end{figure}

\subsection{LyC candidate: {\it Ion2}}

Another source at $z=3.218$ (CANDELS ID 18320, \citealt{guo13}) with a significant excess in the ultraviolet spectrum
has been  identified in the same field 
with a $U$-band magnitude $U=26.85 \pm 0.06$ and 
color ($U-B$)=$1.66 \pm 0.07$.\footnote{The $B$-band data used here come from the ultradeep 
VIMOS observation of GOODS-S, Nonino et al. in preparation.} 
At this redshift the $U$-band probes the wavelength range $\lambda<948$\AA~rest-frame,
and any significant signal suggests a LyC leakage.
From the observed ($U-B$) color and following the methodology described above, 
we derive {\em P(fesc=0)}=0.0\% (Fig.~\ref{CS2}), which confirms
that this source is 
a LyC candidate (named here {\it Ion2}). This result also supports the non-zero emitting 
flux blueward of the Lyman limit 
that was tentatively detected in the VIMOS low-resolution spectrum, $R=180$ 
(see Fig.~\ref{CS1}), but with a low $S/N\lesssim2$ \citep{balestra10}.

By examining the morphology of the system in more detail, two 
distinct objects are visible at a separation of $0.2^{\prime\prime}$ in the HST/ACS imaging
(corresponding to a physical separation of 1.5kpc if they are at the same redshift, Fig.~\ref{CS2}), 
both of them showing a drop blueward of the $V_{606}$-band.

In the following, we refer with  a) to the brighter and with b) to the fainter component.
They have been detected as a single object both in the 
optical ACS catalog v2.0 (\citealt{giava04} and see \citealt{vanz09}) 
and in the near-infrared CANDELS catalog \citep{guo13}.
They are not detected in the X-ray 4Ms Chandra observations,
 corresponding to $1\sigma$ limit of $L_{X}<4.1\times10^{42}~erg~s^{-1}$.
That these are two sources further complicates the interpretation. In the following sections 
we investigate the nature of the system in more
detail.

\subsection{Photometric and spectroscopic confirmation of LyC candidates}

\subsubsection{LyC emission}
\begin{figure}
\centering
\includegraphics[width=8.5 cm, angle=0]{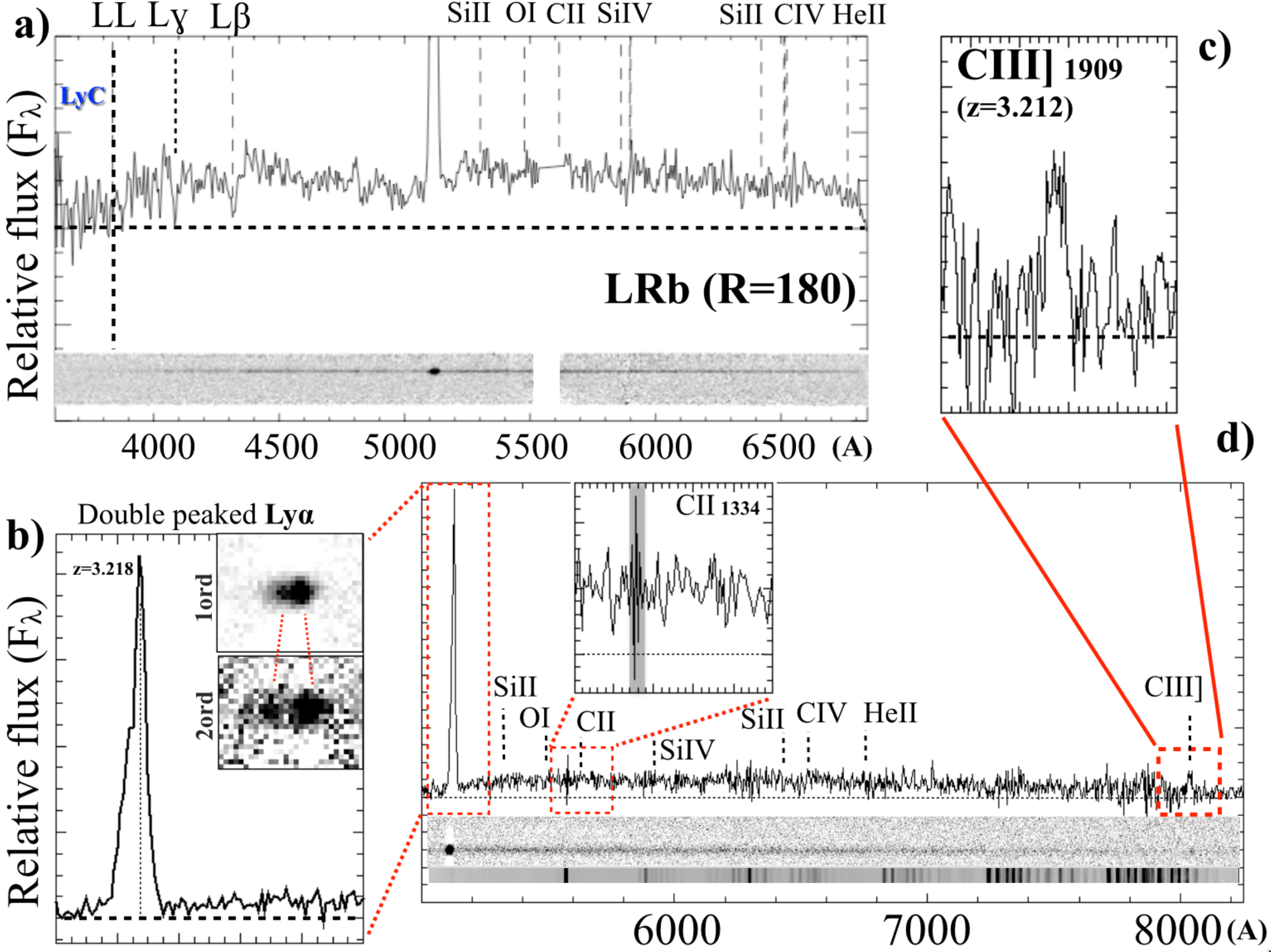}
\caption{Low- ($R=180$) and medium- ($R=580$) resolution VIMOS spectra of {\it Ion2}
in panels a) and b). The two spectra overlap in the range
(4800-6800\AA). Several typical low- and high-ionization atomic transitions are reported. The
$Ly\alpha$ emission in both spectra and the $Ly\beta$ and $Ly\gamma$ absorption
features in the low-resolution spectrum (panel a) are clear. The Lyman continuum blueward of the Lyman limit (LL)
is indicated in panel a). Panel b) (bottom left) shows the zoomed one-dimensional $Ly\alpha$ line
extracted from the medium-resolution spectrum (R=580). Its atypical blue asymmetric tail is clear.
The insets in panel b) show
the two-dimensional $Ly\alpha$ line at first (top) and second (bottom) order. The latter, 
with double spectral resolution, $R\simeq 1200$, clearly shows the double-peaked structure
of the line.
In the top right panel (panel c) a zoom-in of the position of C\,{\sc iii}]$\lambda 1909$ is shown at $z=3.212$.
The insets in the panel d) show the region in which the C\,{\sc ii}]$\lambda 1334$ absorption line would lie:
absorptions consistent with noise fluctuation are present (the gray stripe indicates the position of the
sky emission line, 5577\AA).
}
\label{CS1}
\end{figure}

Another useful feature is detected in the medium-resolution spectrum obtained
with VIMOS ($R=580$) that was published in \citet{balestra10}.
The $Ly\alpha$ line shows a clear and atypical asymmetric
blue tail, at variance with the typically opposite shape toward red wavelengths 
(\citealt{shapley03}; \citealt{ver08}; \citealt{vanz09}). We performed a dedicated reduction of the
VIMOS quadrant containing {\it Ion2} following the reduction prescription
described in \citet{vanz14b}. The two-dimensional spectrum is shown in
Fig.~\ref{CS1} (panel d), in which the C\,{\sc iii}]$\lambda 1909$ 
emission line at $z=3.212$ is clearly identified (S/N=8), and the $Ly\alpha$ was 
detected at the first and second order (the latter with a double spectral resolution, 
$R\simeq1200$). In particular, the second-order detection shows a clear double-peaked emission
with $dv \simeq 550 \pm 90 km~s^{-1}$ (see Fig.~\ref{CS1}, inset of panel b).
There are at least two possible explanations for the two lines in emission around the
$Ly\alpha$ position:
(1) the two sources identified in the ACS images are at a very similar redshift and 
both with $Ly\alpha$ in emission and detected in the MR spectrum; (2) one of them 
has an intrinsic double-peaked $Ly\alpha$ line in emission (as observed, e.g.,
in \citealt{karman14}; \citealt{kulas12}).
It is worth noting that $Ly\beta$ and $Ly\gamma$ lines have been clearly 
detected in absorption in the LRb spectrum; these detections are possible only for the brighter 
of the two components. The fainter component has an 
estimated magnitude at the wavelength interval encompassing the $Ly\beta$ and $Ly\gamma$ 
absorption lines (i.e., $\simeq$ $B_{435}$-band) of $B_{435}=27.25\pm0.24$, which is too faint 
to allow a detection of the continuum in the VIMOS LR spectrum with only four hours of integration 
time (expected to have $S/N \sim 1$, running the ESO/Exposure Time 
Calculator)\footnote{\it http://www.eso.org/observing/etc/}, 
and therefore the detection of any absorption line is prohibitive, too. 
This implies that they are associated with the brighter component (at $z_{Ly\alpha}=3.21$),
that is, two magnitudes brighter in the $B_{435}$-band.
The uncertainty remains whether the (fainter) source (b) is at the same redshift
as the main object or not.
If it is at a very similar redshift, then it would be interesting to also investigate 
its ionizing emission.
Conversely, a lower redshift source $z<<3.2$ would be less interesting and 
could contribute to the observed $U$-band magnitude (but not to the LR spectrum, being too faint).  
The HST/ACS resolution allows us to perform dedicated photometry of both components, 
in particular, components a) and b) were subtracted from each ACS band 
(see Sect.~4.6),  and the photometry was properly derived by running SExtractor.
The two components show very similar photometric behaviors; we derive the following
ACS $BViz$ magnitudes for the (brighter) source a):  $25.34\pm0.13$, $24.72\pm0.05$, $24.66\pm0.02$, and $24.64\pm0.12,$
and for the fainter source b):  $27.25\pm0.24$, $26.63\pm0.09$, $26.41\pm0.04$, and $26.35\pm0.20$. 
 Source b) also shows a clear drop in the $B_{435}$-band,
appreciable in Fig.~\ref{CS2}. If interpreted as due to the IGM 
decrement, it would be compatible with the redshift and the decrement also observed in the 
a) companion ($z_{Ly\alpha}=3.218$). The ($B_{435}-V_{606}$) colors of the two components are practically 
the same, $0.62\pm0.14  $ for a) and $0.62\pm0.26$ for b), while the
$B_{435}-i_{775}$ drop is slightly larger for the fainter one, supporting its high-z nature.
The other possible interpretation for the drop is the Balmer break at $z\simeq0.2$.
However, at this redshift and given 
the spectral coverage in the range $3500-10000$\AA, typical features such as the [O\,{\sc ii}]$\lambda 3727$, 
[O\,{\sc iii}]$\lambda 4959-5007$, $H\beta,$ and $H\alpha$ would be easily detectable in the spectra
if it were an emission line object.
Regardless of the redshift of the fainter component, the estimated
$U$-band magnitude of $26.85\pm0.06$ for the system cannot be explained by the b) source alone,
even assuming that the b) component maintains the same $B_{435}=27.25 \pm 0.24$ magnitude in the $U$-band as well.

The two distinct and close objects, the two lines in emission, 
and the drop blueward of the $V_{606}$-band in both objects
support the interpretation that the two sources are at very similar redshift with a velocity difference 
of $\Delta v \simeq 550 \pm 90~km~s^{-1}$, with the two $Ly\alpha$ lines resolved in 
the $R=1200$ spectrum (they are not spatially resolved with the given
slit orientation).
Under this assumption, the system remains a convincing LyC candidate,  with {\em P(fesc=0)}=0.

\subsubsection{Low gas-covering fraction ?}

\citet{heckman11} suggested a link between the low-ionization [C\,{\sc ii}]$\lambda 1334$ 
(and [Si\,{\sc ii}]$\lambda 1260$) absorption line
and the transparency of the interstellar medium to ionizing photons, 
such that a non-zero residual flux in the [C\,{\sc ii}]$\lambda 1334$ line is a necessary condition 
to have a low optical depth at the Lyman edge (see also \citealt{jaskot14}). 
\citet{borta14} found similar features in a local Lyman-break analog with Lyman-continuum leakage. In the present case, the absence of the low-ionization absorption
lines of silicon and carbon (see Fig.~\ref{CS1}), especially the faintness of the multiple atomic 
transitions (i.e., the evident non-zero residual flux) of silicon [Si\,{\sc ii}]$\lambda 1260, 1304,   1526,$ 
and [C\,{\sc ii}]$\lambda 1334$ suggests a low covering fraction of neutral 
gas (see \citealt{jones13}). The available spectral resolution prevents us from performing a more quantitative study,
but the absence of these lines supports a possible high transmission along the line of sight, 
consistent with the potentially high escaping ionizing radiation we inferred indirectly
from the MC analysis.

As concluded in the previous section, it is plausible that the two components are at the same
redshift, both with $Ly\alpha$ in emission. 
Another piece of information comes from the semi-forbidden doublet (here
  unresolved) nebular C\,{\sc iii}]$\lambda\lambda1906-1909$.  In particular, the $z_{CIII]}=3.212$
and the redshift of the bluer peak of the $Ly\alpha$ structure, $z_{Ly\alpha,blue}=3.211$, 
are fully compatible within an uncertainty of $70km~s^{-1}$ ,
while the red peak is redshifted by about $550 km~s^{-1}$ ($z_{Ly\alpha,red}=3.218\pm0.001$).
If we adopt the C\,{\sc iii}] transition as a proxy of the systemic redshift,
then the emerging $Ly\alpha$ photons at the systemic redshift could further support a scenario of a
low neutral gas column density that would be consistent with the possible ionizing leakage
 (the resonant scattering is reduced and the $Ly\alpha$ 
photons escape not far from the resonance frequency, close to the systemic velocity,
see \citealt{ver14}; \citealt{schaerer11}). 
If this is the case, the escaping ionizing and $Ly\alpha$ photons could share a similar 
(or have the same) physical path through the cavities in the interstellar medium (\citealt{beh14}).

\begin{figure}
\centering
\includegraphics[width=8.5 cm, angle=0]{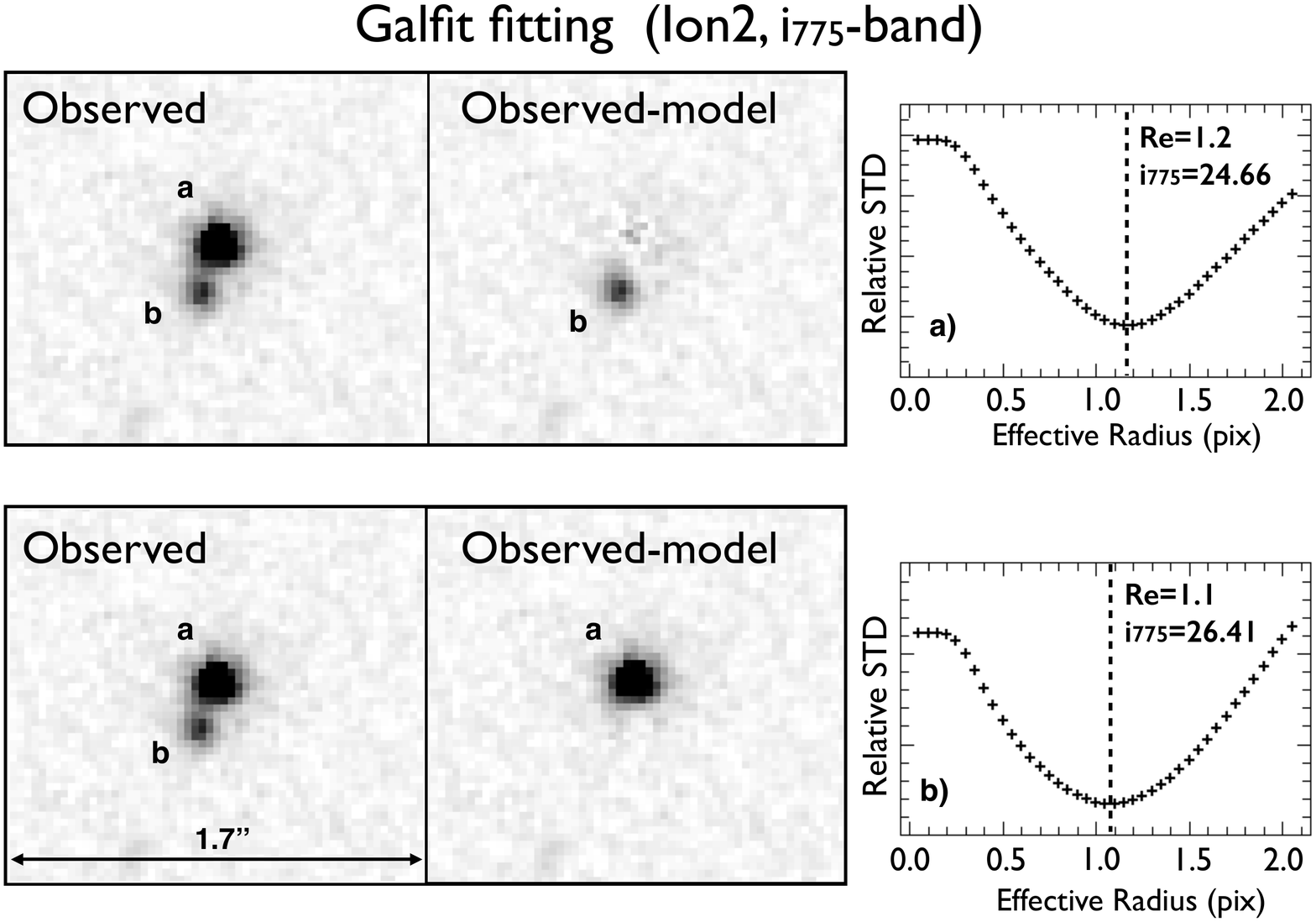} 
\caption{Example of the GALFIT fitting in the $i_{775}$-band of the two components 
of {\it Ion2} . In the top and bottom panels the a) and b) components have been
subtracted. The best solution provides an effective radius $R_{e}\simeq 1.1$
pixels.
}
\label{galfit}
\end{figure}

\subsection{Signature of variability: faint AGN component ?}

The ACS/$B_{435}$ and VLT/VIMOS $B$-band observations have been obtained 
in two epochs, in 2002 and 2010, respectively.
The two filters encompass a very similar wavelength intervals with a small difference
in their shape (see Figs.~\ref{ion1} and ~\ref{CS2}).
We quantified the magnitude difference ($B_{VIMOS}$-$B_{435}$) 
to be in the range $0-0.1$ (and average $+0.03$)
by convolving the two filters with a large control sample made of BC galaxy templates.
The difference of the observed magnitudes $\Delta B$=($B_{VIMOS}-B_{435}$) at the two epochs 
(corrected for the average offset) and the uncertainty $\sigma_{\Delta B}$ can be computed. 
The quantity {\em BVAR}=ABS($\Delta B / \sigma_{\Delta B}$) provides an estimate of 
the significance of the time variation between the two epochs (e.g., \citealt{vill10}). 
The rest-frame interval of time probed  at redshift $\sim$3 is $\simeq$ two years.
As a sanity check, we calculated this quantity for the known AGNs in the GOODS-S field 
(\citealt{szo04}; \citealt{xue11}), and for several of them 
the {\em BVAR} ranges between $6-10$, meaning that the variation is evident and significant at more than
six sigma (as expected for AGNs, especially in the ultraviolet, e.g., \citealt{cristiani97}). 
This will be explored in a forthcoming work on the full photometric sample in the GOODS-S field. 
Here we limit our analysis to the two examples reported above. 
The derived {\em BVAR} for {\it Ion1} and {\it Ion2} is 1.3 and 3.0, respectively
(corresponding to a magnitude difference $\Delta B = +0.26$ and $\Delta B = +0.42$).
The two sources are isolated and not subject to photometric
contamination from other nearby objects.
While for {\it Ion1}  there is no significant evidence for variability,
in the case of {\it Ion2} it is significant at a $3\sigma$ level.
Since the variability is a sufficient condition to confirm nuclear activity, its detection
in the case of {\em Ion2} suggests that an AGN component is highly
likely.
We recall that the two sources are currently not identified with deep 4Ms X-ray observations 
\citep{xue11}, corresponding to $1\sigma$ limit of $L_{X}<4\times10^{42}~erg~s^{-1}$ for both.
In addition, the typical high-ionization spectral features of nuclear activity
(e.g., N\,{\sc V}~$\lambda 1240$,  C\,{\sc iv}~$\lambda 1550$) are absent as well.
However, in a fainter luminosity domain the situation could be different.
\citet{silver10} have found X-ray selected
AGNs at $2<z<3.6$ with optical spectra fully compatible with those of 
typical star-forming galaxies, that is, without high ionization emission lines 
and/or without any line in emission (see also \citealt{civano11}). Without 
the X-ray imaging the AGN component in these systems would not have been 
identified. The UV slopes of these sources is also compatible with the UV 
slope observed here. \citet{vanz10b} also reported on two AGNs 
at $z=3.462$ and $z=3.466$ \citep{balestra10} with ionizing radiation
directly detected in the $U$-band ultradeep VIMOS observations, both 
showing a blue color $(i_{775}-z_{z850}) \lesssim 0$. 
The two sources {\it Ion1} and {\it Ion2} 
are $\sim 1$ magnitude fainter, not detectable in 
the deep X-ray observations, and may represent a regime in which 
the stellar and non-stellar ultraviolet emissions are similar.

\subsection{Spatially resolved emission}
High spatial resolution ACS imaging has been fitted with GALFIT \citep{peng10}
to extract the basic PSF-corrected morphological parameters. The assumed PSF in the HST
images has been derived empirically from observed stars close to the
targets.

First, we note that {\it Ion1} and
{\it Ion2} (both the $a)$ and $b)$ components for the latter) show a compact uncomplicated shape, so a good fit is reached by adopting a 
simple Gaussian profile (S\'ersic model with $n = 0.5$) and leaving the effective 
radius {\em Re}, the axis ratio B/A, the coordinates X,Y, the magnitude, and the 
position angle (PA) as free parameters
(very similar results are also obtained by leaving the S\'ersic index $n$ as a free parameter). B/A and PA are not relevant parameters in the fit, 
since all components have a circular symmetry, therefore we fixed them to B/A=1 and PA=0. 
The coordinates of the sources and their magnitudes were fixed to the best 
guesses derived from SExtractor.
Then GALFIT was run many times by varying the {\em Re} parameter from 0.025 to 2.0 pixels.
An example is shown in Fig.~\ref{galfit}, in which the two components
of {\it Ion2} were modeled and subtracted from the $i_{775}$-band ($\simeq 1800$\AA~rest-frame).
The behavior of the residuals at the position of the source 
is shown as a function of the effective radius (right panel of Fig.~\ref{galfit}).
The minimum of the residuals for $Re\simeq1.1-1.2$ pixels is clear 
(1pix=$0.03^{\prime \prime}$), and the residuals map provided by GALFIT, as a result 
subtracting the best-fit model from the original galaxy, do not show 
significant structures (middle panels of the same figure).

We performed this exercise in all the ACS bands  ($\simeq 1000-2000$\AA~rest-frame), 
both for {\it Ion1} and {\it Ion2}. 
Both sources (and the two subcomponents of {\it Ion2}) are spatially resolved in the 
ultraviolet, with effective radii of 200 and 300 parsecs ($\pm~70$~pc), respectively.
While in the previous section we discussed the possible presence of an
AGN component, especially from the variability of {\em Ion2}, the spatially resolved
emission suggests that stellar UV light also contributes to the emerging spectrum.
Unfortunately, high spatial resolution imaging in the
ionizing domain is not available at the moment. 
Moreover, the compactness of the central star-forming region 
within a few hundred parsecs and the leakage of ionizing photons has 
recently been reported  in a local starburst galaxy \citep{borta14},
in which a high volume density of young stars can generate intense winds
able to propel the surrounding interstellar neutral and ionized gas outward 
at velocities in excess of $1000~kms^{-1}$ \citep{heckman11}. The two sources 
discussed here could be the higher redshift counterparts and will
merit future investigation of the possible presence of winds.

\section{Discussion and conclusions}

The rest-frame far UV colors of star-forming galaxies, which at $z\gtrsim 2.7$
are observed in the optical window, are powerful diagnostics of the forming
stellar populations and of the ISM. They are also very useful tools for selecting the galaxies themselves by means of dedicated color-color
diagrams, such as the Lyman-break technique (\citealt{giava02}), or by means of
photometric redshift selection. The spectral region most often targeted for the
construction of colors typically straddles the 912\AA\ Lyman-continuum
discontinuity and the $Ly\alpha$ resonant line at 1216\AA.
In general, the dispersion of the observed colors results from the
contribution of a number of independent factors, which include the following: 

1) the intrinsic scatter of the UV SED of galaxies. This contrbution primarily
depends on the scatter of the integrated colors of massive stars and on the
scatter of reddening by dust. The former term is dominated by variations of
the age of the starburst and variations of the IMF, the latter by variations
of the total amount of dust (which is commonly parameterized in terms of the
E(B-V) color excess) and of variations of the reddening curve;

2) the scatter of the cosmic opacity, which consists of the Lyman-alpha
forest and of DLAs and LLSs. Detailed analyses have been performed in this
work by running MC simulations based on the prescriptions of I08 and I14;

3) the scatter of the amount of escaping ionizing radiation, which depends on
the HI column density of the ISM of the source along the line of sight. There
are no firm constraints on this term, but its effects on the observed color are
certainly weak because the number of leaking ionizing photons is small
in absolute terms compared to non-ionizing photons.
Another cause is the scatter of broad--band colors constructed
with passbands that sample the spectral regions to the blue and the red of
the Lyman limit are much more strongly affected by small differences in the redshift of
the targeted galaxies, which determine the amount of non-ionizing flux that
enters the blue passband. This amount of flux is generally much larger than the
potential leaking ionizing flux, and so is the corresponding contribution to
the scatter of colors. Nevertheless, knowledge of the exact 
position of the Lyman limit (spectroscopic redshift) has allowed us to 
define a method for identifying the signature of a possible
escaping ionizing radiation;

4) photometric scatter. This term depends on the depth of the photometry used
to construct the colors, and thus on the experimental configuration. It is
usually accurately measured as a function of the luminosity of the source by
means of Monte Carlo simulations. It can provide a comparatively strong
contribution to the scatter, especially in colors that include spectral regions
where the source is faint, for instance, blueward of the Lyman limit, the Lyman-alpha
forest, or both;

5) confusion scatter. This term arises from photometric contamination by
nearby spurious interlopers at lower redshift than the targeted source 
(e.g., \citealt{vanz10a}). It depends on the angular resolution of the images
used to construct the colors, and thus on the instrumental configuration and
ground-based seeing (if applicable). Even if the interloper is a very faint
galaxy, the contamination can be severe in colors that include spectral regions
where the source is faint, for example, blueward of the Lyman limit, the Lyman-alpha 
forest, or both.

We analyzed the effect of an escaping ionizing radiation on color selection criteria by running dedicated Monte Carlo simulations. 
On the basis of similar simulations anchored to the observations, a method of selecting 
LyC continuum emitter candidates from multiband photometry was also discussed and applied to a sample of galaxies.

These are the main results:

\begin{enumerate}
\item{The redshift evolution of the mean free path of ionizing photons 
   through the intergalactic medium and a proper treatment
    of the ionizing continuum (galaxy template) indicate that
    there is no significant effect of LyC leakage on classical 
    broad-band photometric selection techniques (e.g.,$U_{n}GR$ or $BVI$), 
    at variance with what has been reported by \citet{cooke14}. The effect of an
    ionizing emission is more significant as the width of the band
    encompassing the Lyman edge decreases and approaches 
    the expected mean free path at the given redshift (being maximized 
    in the narrow-band surveys or deep spectroscopy, e.g., \citealt{mostardi13}; 
    \citealt{giallongo02}; \citealt{shapley06}). 
    Under the assumption that the IGM prescription of
    I14 is a good representation of the IGM structure,
    no significant deviation was found in the non-ionizing part of the
    IGM attenuation ($912-1216$\AA) either, that is,   the scatter is not larger 
    than 0.4 mags. 
    Overall, the median color tracks do not differ significantly 
    if {\em fesc} = 0 or 1, even in the most extreme stellar ionizing emitters 
    (age=1Myr, E(B-V)=0, $Z/Z_{\odot}=0.02$).}

\item{A method for photometrically selecting LyC-emitter candidates 
    was described and applied to a sample of 35 galaxies belonging to the
    GOODS-S field. Two candidate LyC emitters were identified:
    one known LyC emitter at $z=3.795$ was successfully recovered
    ({\it Ion1}). Another source with possible LyC leakage
    was reported ({\it Ion2}). The probability of a null escaping ionizing radiation
    is low ($\le 5\%$) for both sources. Detailed analysis based on
    morphology, variability, and the ultraviolet excess suggests that a
    probable contribution from faint AGN activity could dominate 
    the LyC emission and/or have played a crucial role in making the medium 
    transparent. The spatially resolved stellar UV emission 
    also suggest there could be a contribution by stellar
    radiation.}

\end{enumerate}

    There are caveats for the method described in this work. 
    First, the probability {\em P(fesc=0)} depends on the fraction
    of the band used to probe the LyC region. A large {\em fesc} could be present, for example, but might be missed 
    because only a small fraction of the band probes it.  This effect can be
    attenuated by including bluer overlapping bands that ensure a continuous 
    redshift coverage (as is the case in the typical multifrequency surveys such as GOODS/CANDELS).
    Second, the method tends to select relatively strong LyC emitters on average (it depends on 
    the width of the filters used and the presence of LLSs/DLAs IGM along the line of sight) 
    because they are the most effective in 
    the broad-band contribution. Third, the photometric accuracy is
    crucial in these analyses to identify the position in the color-color 
    plane with good detail and determine $P(fesc)$. The higher the photometric S/N ratio in the relevant
    bands, the more accurate the identification of LyC candidates with this method.
    The interplay among these limitations is included in the
    performed MC simulations, however.

    Despite these limitations, we have demonstrated that 
    regardless of the nature of the source (AGNs, LyC-galaxy, spurious detection, etc.),
    this type of objects can  in principle be automatically identified with this technique.
    The color excess in the UV coupled with morphological information has 
    previously been used to select QSOs/AGNs
    (e.g., \citealt{gabbane07}; \citealt{casey08}; \citealt{salvato09}) and is conceptually 
    similar to what we discussed here. However, the analyses performed here
    are based on the detailed IGM stochasticity and focused mainly on the LyC 
    part of the spectrum.

From point (1) above it follows that the broad-band photometric 
selection of high-redshift galaxies does not prevent selecting galaxies
with LyC leakage.
Nonetheless, a clear sample of ionizing galaxies has not been
identified yet. One reason could be that
the stellar ionizers exist in a much fainter and still unexplored
luminosity domain, 
while relatively bright ($L>0.5L^{\star}$) ionizing sources are rare
and require efficient physical mechanisms to increase the 
transparency of the interstellar medium and favor ionizing photons to
escape. 
The variability detected in {\em Ion2} and the difficulty in reproducing (with
galaxy templates) the observed $U$-band magnitude of {\em Ion1 } ($\lambda < 830$\AA) 
given its observed ultraviolet slope suggest that the  two sources could host a faint AGN that possibly
contributes to the LyC emission. X-ray detected ($L_{X}>10^{42.5}~erg~s^{-1}$) AGNs with blue 
UV-slope and galaxy-like spectra have been reported in the literature 
(e.g., \citealt{civano11}, \citealt{silver10}), and the two sources reported here 
could be AGNs at fainter luminosities (e.g., \citealt{maoz07}).
In other words, while the more massive systems would need an efficient feedback 
(possibly provided by an AGN) to clean the medium and decrease the column density 
of the neutral gas, 
the fainter and still unexplored sources could regulate the opacity of the interstellar
medium with supernova feedback only if they indeed contain a large portion
of the LyC emitter (\citealt{wise14}; \citealt{kimm14}).

In addition to the possible presence of a faint AGN component, the 
spatially resolved emission in the ultraviolet 1000-2000\AA~rest-frame 
suggests that the stellar emission contributes in this wavelength domain.
It is beyond the aim of this work to decompose the two components, 
we only note that the two might also coexist in the LyC domain.
In particular, stellar ionizing photons could share a very similar path
of the interstellar and circum galactic media of these photons arising from the nucleus,
making the system a sort of hybrid-ionizer (stellar and non-stellar). In this respect, high spatial
resolution imaging with HST (WFC3 UV channels) of confirmed LyC emitters would provide precious 
information about the spatial distribution and the relative balance of the two sources
of ionizing radiation.

If there were a faint AGN component in this type of sources,
it might also explain the recently observed extreme ultraviolet flux ratios between ionizing and
non-ionizing wavelengths in various Lyman alpha emitters reported in the literature
(e.g., $f_{1500}/f_{900} < 1-2$), e.g., \citet{nestor13}, \citet{mostardi13}. 
For example, {\em Ion2} is a Lyman alpha emitter without any typical feature, which suggests that it is 
an AGN (from X-ray and spectral features), but its variability suggests that there is an AGN component.
We therefore remark that it is important to investigate the AGN nature at faint luminosity 
limits in this type of studies.

Finally, convolving the stochasticity of the IGM 
with galaxy templates anchored to the observed UV slopes 
and comparing this with the observed colors proved to be an 
efficient automatic tool to determine the reliability of spectroscopic 
redshift measurements or to identify photometric contamination from lower-z 
sources in the UV. In these cases, an addiitonal flux in the ultraviolet is present and
is therefore recovered as a possible ionizing leakage.
This is particularly useful for large spectroscopic redshift surveys
coupled with deep and high spatial resolution imaging, such as VUDS (\citealt{olf15}) 
and/or the VANDELS survey (ESO public survey 
\footnote{\it www.eso.org/sci/observing/PublicSurveys/sciencePublicSurveys.html}), which 
cover the CANDELS fields. We defer the application to 
these large datasets to a forthcoming work.
 
\begin{acknowledgements}
    We would like to thank the anonymous referee for constructive comments
    and suggestions. We thank G. Becker, J. Bolton, V. D'Odorico, and 
    S. Cristiani for useful discussions about the IGM transmission and its 
    variance. E.V. thanks M. Brusa and R. Gilli for stimulating discussions
    and I. Balestra for providing useful information about the VIMOS 
    spectroscopy of Ion2. We acknowledge the financial contribution from the 
    PRIN-INAF 2012. AKI is supported by JSPS KAKENHI Grant Number 26287034.
\end{acknowledgements}


\end{document}